\documentstyle[psfig,12pt]{article}
\def\be{\begin{equation}}
\def\ee{\end{equation}}
\def\bea{\begin{eqnarray}}
\def\eea{\end{eqnarray}}
\def\br{\bf r}
\begin{document}
\input epsf
\setcounter{page}{1}
%\begin{frontmatter}
\baselineskip 4ex

\centerline{\large \bf Exotic Structure of Carbon Isotopes}
\vspace{1.cm} 
\centerline{Toshio Suzuki}  
\centerline{Department of Physics, College of Humanities and
 Sciences, Nihon University}     
\centerline{Sakurajosui 3-25-40, Setagaya-ku, Tokyo 156-8550, Japan}
\vspace{0.3cm}
\centerline{and}
\centerline{H. Sagawa}
\centerline{Center for Mathematical Sciences, the University of Aizu}    
\centerline{Aizu-Wakamatsu, Fukushima 965-8560,  Japan}  
\vspace{0.3cm}
\centerline{and}
\centerline{Kouichi Hagino}
\centerline{Yukawa Institute for Theoretical Physics, Kyoto University}
\centerline{Kyoto 606-8502, Japan}
\vspace{1.0cm}

\begin{abstract}
We  studied firstly the ground state properties of  C-isotopes 
using a  deformed Hartree-Fock (HF)+ BCS model  with Skyrme interactions. 
Shallow deformation minima are  found in several neutron$-$rich 
C-isotopes.  It is shown also that 
the deformation minima appear in both the  oblate and the prolate sides
in  $^{17}$C and  $^{19}$C having almost the same binding energies.
Secondly, we  carried out shell model calculations to study electromagnetic 
moments and electric dipole transitions of the C-isotopes.  We point out 
the clear configuration dependence of
the quadrupole and magnetic moments in the  odd C-isotopes, which will 
be useful to find out the deformations and the spin-parities of the 
ground states of these nuclei.  
We studied  electric  dipole states of C-isotopes 
focusing on the interplay between low energy Pigmy strength and  giant 
dipole resonances.   
 Reasonable agreement is  obtained 
with available experimental data for 
the photoreaction cross sections 
both in the low energy region below $\hbar \omega $=14 MeV and in the high
energy giant resonance region (14 MeV  $<\hbar \omega \leq $30
MeV).  
 The calculated transition strength below dipole giant
resonance ($\hbar \omega \leq $14 MeV)  in 
heavier C-isotopes than  $^{15}$C  is found to exhaust 
about $12\sim16\%$ of the classical Thomas-Reiche-Kuhn sum rule value and
$50\sim80\%$ of the cluster sum rule value. 

\vspace{0.5cm}

%\noindent
\centerline{PACS numbers/21.60.Cs, 24.30.Cz, 25.20.-x}
%\centerline{keywords: quadrupole moment, magnetic moment,
%  , giant dipole state, pigmy state}  
\end{abstract}
%\end{frontmatter}
%\bigskip

\newpage

%\vspace*{0.6cm}
%\normalsize\baselineskip=22pt
%\setcounter{footnote}{0}
%\renewcommand{\thefootnote}{\alph{footnote}}

\section{INTRODUCTION}

The structure of  nuclei far from the  $\beta-$stability lines 
is often very different from that of stable nuclei 
due to largely extended wave functions as well as the large asymmetry 
between neutron and proton mean fields. Because of  
these unique features of the mean field, 
a naive 
extrapolation of the $\beta-$stable nuclei will fail to 
predict the structure of these exotic nuclei.  A  
typical example is the new shell structure at the neutron number 
$N=$16 in O isotopes \cite{Ozawa}.  
The structure of dipole excitations 
in neutron-rich O isotopes is 
also found out very different 
from that of stable nuclei, especially in the low energy region
below GDR\cite{HSZ98,Emi98,SS99}. 

A study of heavy C-isotopes is another current issue, 
where these exotic structures 
may be expected. 
In this paper, we study the ground  state properties 
   of C isotopes 
performing deformed Hartree-Fock (HF) +BCS calculations and
also shell model calculations. The energy surfaces of 
C isotopes are studied as a 
function of the quadrupole deformation parameter $\beta_2 $
in order to find out the deformation minimum for different
single-particle configurations. 
Special emphasis will be put on 
the magnetic and quadrupole moments (Q-moments) for odd C isotopes
which will manifest their exotic structure.
Electric dipole excitations of C isotopes are investigated 
by a large scale
shell model calculations focusing on the interplay between 
low energy Pigmy strength and  giant 
dipole resonance (GDR).  We try to find out  
the deformation effect on the dipole strength distributions, 
which will  increase the width of GDR.

The paper is organized as follows. 
In section 2, we present results of the deformed Skyrme HF +BCS
calculations. The magnetic and Q-  moments are 
discussed in section 3.  The Pigmy and GDR are shown in section 4.  
A  summary and conclusions are given in section 5.

\section{DEFORMED SKYRME HF CALCULATIONS}

In this section, we investigate  the neutron number dependence of
deformation properties  along the chain of  C-isotopes.
For this purpose, 
we perform deformed HF+BCS calculations with Skyrme interactions
SkI4 \cite{RF95}, SIII \cite{S3} and SkM$^*$ \cite{SkM}, 
using the computer code {\tt SKYAX} \cite{PG}.
The axial symmetry is assumed  for  the HF deformed potential.
The pairing interaction is treated in the BCS approximation and is 
taken to be a volume-type $\delta$-interaction 
\be
V({\br_1},{\br_2})=V_0\,\delta(\br_1-\br_2), 
\label{eq:pair_v}
\ee
where the pairing strength $V_0$ is taken to be $-323$
MeV$\cdot$fm$^3$ for neutron and $-310$ MeV$\cdot$fm$^3$ for proton 
\cite{BRRM00}. These values are determined so as to fit the
experimental pairing gaps for several isotope as well as isotone chains 
of semi-magic nuclei, see Ref. \cite{BRRM00} for details. 
The HF+BCS calculations are also  carried out  with a 
density dependent pairing interaction
\be
V({\br_1},{\br_2}) =V_0' \left(1-\frac{\rho (r)}{\rho _0}\right)\,
\delta(\br_1-\br_2)
\label{eq:pair_s}
\ee
where $\rho (r)$ is the HF density at ${\br} = ({\br}_1+{\br}_2)/2$ 
and $\rho _0$ is  chosen  to be 
0.16 fm$^{-3}$.  The pairing strength $V_0^{'}$ is taken to be
$-999$ MeV$\cdot$fm$^3$ for neutrons and $-1146$ MeV$\cdot$fm$^3$ for
protons \cite{BRRM00}. A smooth energy cut-off is employed in the BCS
calculations \cite{BRRM00}.

%Figure \ref{fig:c_def} 
Figure 1 shows the binding energy surfaces 
for even-mass C isotopes 
as a function of the quadrupole deformation parameter $\beta_2$ 
obtained with the SkI4 interaction  together with  
the two pairing interactions. 
In general, both of the two pairing interactions give a similar 
energy surface which is
flat in a wide range of the deformation parameter $\beta_2$. 
The energy minima are tabulated in 
%Table \ref{table:def_mini}.  
Table 1.
For the volume pairing, 
the energy surface of $^{12}$C is rather flat between 
$-0.3 < \beta_2 < 0.3 $. The energy minimum becomes apparent for $^{14}$C.
For heavier C isotopes $^{16}$C and $^{18}$C, two shallow minima appear 
both in the prolate and oblate sides.  In $^{18}$C, the ground state 
has the largest deformation at $\beta_2 $=0.38, while the local minimum 
 appears at the oblate side at  $\beta_2 \sim -$0.3. 
In the case of the surface pairing (eq. (\ref{eq:pair_s})), clear minima are
not seen in the energy surface except for $^{14}$C, although there are 
large flat plateaus between $-0.3 < \beta_2 < 0.3 $ as in
the case of the volume pairing shown in the upper panel of 
%Fig.  \ref{fig:c_def}. 
Fig. 1.

The surface pairing tends to yield a larger pairing gap 
for C isotopes than the volume pairing (see Table 1). In order to
asses whether the slight difference of the energy surfaces obtained 
with the two pairing interaction is due to the different form of 
interaction or not, we 
repeat the same calculations for the surface pairing interaction 
but by reducing the strength $V_0'$ by a half.  Even smaller 
pairing strength is adopted in the study of O  isotopes recently 
in ref. \cite{Khan}.
Figure 2 shows the energy surfaces thus obtained. 
As we see, clear minima now appear for $^{14}$C and $^{16}$C,
suggesting that the energy surface is sensitive to the strength 
of the pairing interaction. 

The results with two different Skyrme interactions 
SIII and SkM$^*$ together with the surface pairing are 
shown in Fig. 3. 
Although the strengths of the pairing interaction may be different 
for each set of Skyrme interactions, we  use  the same values as those 
used with the SkI4 set. 
The energy surface for the C-isotopes show no clear minima, 
but rather flat in the deformation region $-0.3< \beta_2 <0.3$.  
These features are similar as those of SkI4 interaction.

The one quasi-particle state energies obtained with the SkI4
interaction are shown in 
%Figs. \ref{fig:c13},   
% \ref{fig:c15},  \ref{fig:c17} and \ref{fig:c19} for 
Figs. 4, 5, 6 and 7 for
$^{13}$C, $^{15}$C,$^{17}$C and $^{19}$C, respectively.
The odd nuclei $^{13,15,17,19}$C are  treated as one 
quasi-particle state on top of the BCS ground state of
neighboring even nuclei.
The Pauli blocking effect of the  valence  particle is not taken into 
account in the present calculations.  
For $^{13}$C, 
the $1/2^- $ state shows  the deepest spherical minimum at $\beta_2 = 0.0$, 
while the two minima are seen in the 1/2$^+$ state at the oblate and the
prolate deformations.  The 5/2$^+$ state has one minimum at around 
 $\beta_2 = -0.36$. The two pairing interactions 
(\ref{eq:pair_v}) and (\ref{eq:pair_s}) give essentially the same results 
for  $^{13}$C.  
These energy minima for the different configurations 
are expected from the Nilsson diagram of deformed harmonic oscillator 
potential\cite{BM75a}.  The  results of $^{15}$C are shown in   
%Fig. \ref{fig:c15}.
Fig. 5. 
The 1/2$^{+}$ state is the lowest
at the prolate deformation 
with $ \beta_2 \sim $0.2 with the volume pairing and  $ \beta_2 \sim $0.12 
with the surface pairing.  There is also a local minimum for the 1/2$^+$ 
configuration in the oblate side.  The minimum of 
5/2$^+$ state appears at the oblate side and the energy 
is 2MeV above that of 1/2$^+$ state.
The three configurations 1/2$^+$, 3/2$^+$ and 5/2$^+$ are competing 
in $^{17}$C as shown in 
%Fig. \ref{fig:c17}. 
Fig. 6.
In the case of the volume 
pairing,  the  3/2$^+$ state has the lowest minimum at the prolate 
deformation $ \beta_2 \sim $0.4 and another local minimum is also found 
at the oblate side with $ \beta_2 \sim - $0.2.  The energy surface of 
the 1/2$^+$ state is similar to that of  3/2$^+$ state although the 
oblate minimum is lower than the prolate minimum in the case of 
the 1/2$^+$ state.
The 5/2$^+$ state has a minimum at the oblate deformation.  
The minima of the
three configurations at the oblate side show very similar 
 $ \beta_2 $ values having almost the same  binding energies. 
 Among all minima, the lowest one is found at the prolate
deformation at $ \beta_2 \sim $0.4 for the  3/2$^+$ state. 
The binding energies of the three states are also very close 
in the shell model calculations of $^{17}$C,  as we discuss in the 
next section.  
The results of $^{19}$C are shown in 
%Fig.  \ref{fig:c19}.  
Fig. 7.
The competition of the three configurations 1/2$^+$, 3/2$^+$ and
5/2$^+$ is apparent as in $^{17}$C. 
The one quasi-particle 
states show the lowest minimum for the 3/2$^+$ state at the oblate 
deformation  $ \beta_2 \sim -$0.3, while the 1/2$^+$ state show the minimum 
at the prolate deformation  $ \beta_2 \sim $0.32.  The 5/2$^+$ state does not 
show any clear minimum as the one quasi-particle state.

In general, the prolate deformation occurs at the beginning of the shell
while the oblate deformation occurs at the end of the shell.
 Experimentally, 
$^{12}$C and $^{13}$C are known to have oblate deformation\cite{Wag,Asp}, 
and $^{14}$C
with $N$=8 becomes almost spherical at the neutron shell closure.
In the present calculations, the energy surface of $^{12}$C is rather
flat between $-0.3 < \beta_{2} < 0.3$. 
 They also  show that C-isotopes with $N$=9$\sim$11,  i.e.,
 $^{15}$C
$\sim ^{17}$C, favor prolate deformation, which is natural as the new
shell begins to be occupied after $N$=8.
 The $^{19}$C nucleus with $N$=13 is shown most likely to favor 
oblate deformation.
This suggests that the neutron number $N$=13 might locate 
in the latter half between the two closed shells and 
could be a manifestation of the new shell closure at $N$=16 instead of
$N$=20 as in the case of O-isotopes\cite{Ozawa,Colo01}. 
 It is thus desperately desired 
 to have decisive experimental information
on the signs of the deformations in heavier C isotopes. 

The neutron number dependence of the deformation in C isotopes was 
studied by using anti-symmetrized molecular dynamics (AMD) model in ref.
\cite{Enyo}.  They pointed out similar neutron number dependence for the 
deformation to the present results in neutron-rich C isotopes.
Namely the neutron deformation changes from spherical in $N$=8, to prolate 
in $N$=10 and then, to oblate in $N$=14, 
while the proton deformation stays always 
oblate independent to the neutron number.  There is a difference in the case 
of   $N$=16.  The present deformed HF+BCS result shows a spherical minimum 
for  $^{22}$C, while the AMD model gives a triaxial shape for the neutron 
configuration.  Since the present deformed HF+BCS model is 
performed assuming the axial symmetric 
 deformation, the two results are not completely equivalent.
It might be interesting to study further the deformation changes taking into
account the tri-axial degree of freedom in the deformed HF+BCS
model. 

\section{MAGNETIC AND QUADRUPOLE MOMENTS AND EFFECTIVE OPERATORS}
 
 We next  perform the shell model calculations  for  
C isotopes with the 
effective interactions WBP10  in the  (0p-0d1s)
configuration  space\cite{WB91} to study the magnetic and the Q- 
moments.
The WBP10 interaction is designed to reproduce  systematically 
the energy of 
ground state and excited states of stable sd  shell 
nuclei.  The energies  and the spin-parities of the states 
near the ground states of odd C isotopes are tabulated in 
%Table \ref{tabel:energy}.
Table 2.
It is interesting to see that two or three different spin states 
are almost degenerate in the odd C isotopes  $^{15}$C, 
$^{17}$C and $^{19}$C.
These degeneracies are also expected from the results of 
deformed HF calculations,  as we showed in the previous
section. 
Several experimental efforts have been  made  to assign   
the spin-parities of odd C- isotopes $^{15}$C, $^{17}$C and $^{19}$C. 
For this end, the magnetic moments and Q-moments will provide 
the most conclusive information.  Calculated magnetic moments and
Q-moments are given in 
%Table \ref{table:moments}.
Table 2.
The effective spin $g_s -$factor is taken to be 0.9$g_s $(bare) 
for neutrons.  This  quenching factor is somewhat larger than the commonly 
adopted values 0.7$-$0.8 in stable nuclei.  This difference might be 
due to smaller effect of the second-order effects in the neutron-rich 
light nuclei
\cite{Suzuki_ex}. 
In the ground state of $^{15}$C, the calculated $g$ factor is
 $-$3.37$\mu _N $ which agrees well with the experimental one 
$|g|$=3.440$\pm $0.018\cite{C15mm}. 
  The calculated values for  the  3/2$^+ $ and 
5/2 $^+ $ states of $^{17}$C are close to be $-$0.514$\mu _N $ 
and $-$0.505$\mu _N $, 
while that of the 1/2$^{+}$ state is $-$2.82$\mu _N $.  The empirical 
value $|g$(exp)$|$=0.5054$\pm $0.0025 \cite{C17mm} 
 excludes the  1/2$^{+}$ state from 
the ground state candidate, while the other two 3/2$^+ $ and 
5/2 $^+ $  states show good agreement within a few percent accuracy.  
There is a complemental experimental data of the
 selection rule on  $\beta $ decay from $^{17}$C to $^{17}$N which favors
the spin 3/2$^+ $   as the ground state of  $^{17}$C.

The  effective charges $e_{eff}(E2) $ are commonly adopted for the shell model 
calculations of Q-moments because of the limitation of the model space.
In ref. \cite{BM75b},  the polarization charges $e_{pol} (E2) $  for the
electric quadrupole moment are calculated by the 
harmonic vibration model and the isospin dependence is given by
\bea
 e_{pol} (E2)/e &=& e_{eff} (E2)/e - \frac{1}{2}(1-\tau_z)   \nonumber\\
    &=& \frac{Z}{A} \chi (\tau =0) + \frac{1}{2} \chi (\tau =1)\frac{N-Z}{A}
% \nonumber\\
     + (- \frac{1}{2} \chi (\tau =1)
   + \frac{Z}{A} \chi (\tau =0)\frac{V_{IV}}{4V_{IS}}
           \frac{N-Z}{A})\tau _z  \nonumber\\
\label{eq:e2_pol}
\eea    
where $ \chi (\tau =0)$ and  $ \chi (\tau =1)$ are the isoscalar  $ (IS)$ and 
 the  isovector $ (IV)$ polarizability coefficients and  
$\frac{V_{IV}}{V_{IS}}$ is the ratio of  $ IV $ and
 $ IS $ components in the static nuclear potential.  The  polarizability
   coefficients are  evaluated to be
 $ \chi (\tau =0)=$1.0  and  $ \chi (\tau =1)=-$0.64 by the 
harmonic vibration model and  the ratio $\frac{V_{IV}}{V_{IS}}$ is taken to be
  $\frac{V_{IV}}{V_{IS}}=-$2.6 from the empirical 
mean field potential strength.  By substituting these values in 
 Eq. (\ref{eq:e2_pol}), we obtain  
\be
 e_{pol} (E2)/e  =\frac{Z}{A}-0.32\frac{N-Z}{A} +(0.32 - 0.65\frac{Z}{A}
  \frac{N-Z}{A})\tau _z
\ee
In ref. \cite{SA01}, a microscopic particle-vibration model  was applied to
calculate the polarization charges in C isotopes  using  HF and random 
phase approximations (RPA).  This model gives state-dependent and isospin
 dependent  polarizability coefficients.  The averaged 
  polarizability coefficients 
  $\bar{ \chi} (\tau =0)$=0.82 and   $ \bar{\chi} (\tau =1)=-$0.24 are found
in $^{12}$C to be smaller than those of the harmonic vibration model.
The smaller  $ \bar{\chi} (\tau =0)$ is mainly due to spreading of IS 
giant quadrupole resonances (GQR), while a large quenching on 
 the $ \bar{\chi} (\tau =1)$  comes from a substantially small ratio 
$\frac{V_{IV}}{V_{IS}}$ of Skyrme interactions compared with
the harmonic vibration model.    
The averaged  polarization charges in active 
valence configurations are  given   by  $\bar{e}_{pol} (E2; n)/e =$
0.53, 0.33 and 0.15 for neutrons  and  
$\bar{e}_{pol} (E2; p)/e$ =0.29, 0.16 and 0.05  for protons 
in $^{12}$C, $^{16}$C  and $^{20}$C, respectively. 
In $^{12}$C, the calculated  $\bar{e}_{pol}$ values give 
 $e_{eff} (n)$=0.53 for neutrons and  $e_{eff} (p)$=1.29 for protons, 
 which agree well
with the commonly used values $e_{eff}$(p)=1.3 and $e_{eff}$(n)=0.5 for 
light nuclei.
The isospin dependence of the  polarization charges might be 
 papametrized from 
the values $^{12}$C and $^{16}$C to be
\be
 e_{pol} (E2)/e  = a\frac{Z}{A}+b\frac{N-Z}{A}+(c+d\frac{Z}{A}
  \frac{N-Z}{A})\tau _z
\ee
with
\be
\,\,\,\,\,\,\,\,\,a=0.82, \,\,\,\,\,b=-0.25,\,\,\,\,\,c=0.12,\,\,\,\,\,d=-0.25.
\label{eq:pc_para}
\ee
Eq. (\ref{eq:pc_para}) gives  $e_{pol)}(E2;n)/e$=0.22 and 
 $e_{pol} (E2; p)/e$ =0.07 for  $^{20}$C. A large difference 
 between the calculated value and that from Eq. (\ref{eq:pc_para}) is 
due to the effect of neutron skin in  $^{20}$C.  The small 
polarization charges in the very neutron rich nuclei are found to be
important to explain the observed Q-moments of B-isotopes\cite{SA01}.

 We use the isospin dependent 
polarization charges in ref. \cite{SA01} to calculate 
Q-moment of C isotopes in Table 2.  In  $^{17}$C, the magnetic $g-$factors 
are essentially the same for the two configurations  3/2$^+ $ and 
5/2 $^+ $   and the calculated g-factors 
 are close to the experimental value.
The calculated Q-moments, however, are very different in the two 
configurations in magnitude and even in sign reflecting the different 
deformation of the two configurations. The neutron and the proton 
contributions for the  Q-moment
   are $17.1 mb$ and $ 6.4 mb$, respectively,  in
  the 3/2$^+ $state, 
while they are   $-5.5 mb$ and $ -3.8 mb$, respectively, 
  in the  5/2$^+ $ state.  
 It should be noticed 
that the magnetic moment and Q-moment of the  
3/2$^+ $ state in $^{17}$C show large deviations
from the single particle values ,$g$(Schmidt) and, Q(s.p.)=
$-37.7 e_{eff}(n)  mb$ for 0d$_{3/2}$ state, i.e., 
even the sign of these moments are different
in the two calculations.  
The  single particle Q-moment for the 0d$_{5/2}^+ $ state is 
Q(s.p.)=$-53.9e_{eff}(n) mb$, which is much larger than the shell
 model prediction Q=$-9.3 mb$ ( Notice the standard value for $e_{eff}(n)
  $=0.5 and 
the presently adopted  value
  for  $^{17}$C is  $\bar{e}_{eff}(n) $=0.33).  However, 
the shell model Q-momemts are
 consistent with the prolate deformation for the  3/2$^+ $state
and the oblate deformation for the  5/2$^+ $ state which are 
suggested by 
the deformed HF 
calculations in Table 1.
Thus the measurement of  Q-moment will be the most decisive 
experiment to assign the spin and the parity of the
ground state of  $^{17}$C and will  provide experimental 
justification of the
deformed HF+BCS and shell model predictions. 
 
The magnetic moments  and the Q-moment of $^{19}$C are given in 
%Table \ref{table:moments}.
Table 2.  It is still under dispute whether the spin and the parity of 
the ground state of  $^{19}$C is 1/2$^+ $ or 5/2$^+ $. 
According to the shell model calculations,  the lowest 3/2$^+ $ is also 
close to the lowest  1/2$^+ $ and  5/2$^+ $ states in energy.
The  neutron and the proton 
contributions to the  Q-moments are $-18.4 mb$ and $ -14.7 mb$ in
  the 3/2$^+ $state, 
while they are   $-0.6 mb$ for neutrons  and $ 1.6 mb$ for
protons  in the  5/2$^+ $ state of  $^{19}$C. 
The values for the 3/2$^+ $state is consistent with the possible 
oblate deformation suggested in Table 1. On the other hand, 
the proton and neutron contributions have different signs in 
the  5/2$^+ $ state,  and also very 
different from the single particle value for the 0d$_{5/2}^+ $ state.
This shell model results suggest  the large configuration mixing in the
 lowest  5/2$^+ $ state of  $^{19}$C.  
It is seen clearly from Table 2 that the magnetic 
moment  and Q-moment   are very different for each configuration 
in $^{19}$C  and will give decisive information on  
 the spin assignment of the ground state of $^{19}$C.

\section{GIANT AND PIGMY RESONANCES IN C ISOTOPES}

 The isovector (IV) 
giant dipole resonance (GDR) is the most well established collective
mode  throughout the mass table with large 
photoabsorption cross sections, exhausting most 
of the classical
Thomas-Reich-Kuhn (TRK) sum rule (the energy
 weighted sum rule value) \cite{Ber75,Wood79,Pyw85}.
As a microscopic
model, we  perform the shell model calculations 
for the  dipole excitation mode in C-isotopes.  The 
calculations take into account 
 a model space of up to 
 (1+3)$\hbar \omega$ excitations  in $^{12}$C, $^{13}$C and  $^{14}$C 
including 0s-0p-1s0d-1p0f shells. Other nuclei are studied in a model 
space of 1 $\hbar \omega$ excitation of 0s-0p-1s0d-1p0f shells.
  The  Warburton-Brown interaction 
WBP10\cite{WB91} is used in this study with  the model space (0s-0p-1s0d-1p0f).
The center of mass spurious components in the wave functions are
pushed up to higher excitation energies by adding a fictitious
hamiltonian which acts only on the center of mass
excitation\cite{Lawson}. 
In a restricted model space, there still remain some spurious
components in the wave functions after the diagonalization of the
model hamiltonian, especially when one uses 
HF or Woods-Saxon single-particle wave functions, instead of harmonic 
oscillator wave functions.  In
order to remove the effect of these spurious components on the
transition strength, 
  we use
the effective transition operator
\begin{equation}
\hat{O} ^{\lambda =1}_{\mu }=e\sum_{i}^{A} (t_{zi}-\frac{N-Z}{2A})r_{i}Y_{1\mu
  }(\hat{r}_i )=e\frac{Z}{A}\sum_{i}^{N}r_{i}Y_{1\mu
  }(\hat{r}_i )-e\frac{N}{A}\sum_{i}^{Z}r_{i}Y_{1\mu
  }(\hat{r}_i )
\label{eq:e1_ope}
\end{equation}
in which  the center-of-mass correction is subtracted from the IV
dipole transition operator.
The transition strength B(E1) is defined  as 
\begin{equation}
B(E1; \omega _n)=\sum_{\mu } \mid \langle n|\hat{O} ^{\lambda =1}_{\mu
}|gs \rangle\mid ^2
\label{eq:be1}
\end{equation}
where the matrix element is calculated between the ground state ($|gs\rangle$)
and the n-th excited 1$^- $ shell model state ($|n\rangle$) with the
excitation energy $\hbar \omega _n $.  
In order to smooth out the discrete strength, 
the transition strength is averaged by a weight factor $\rho (\omega
)$ as
\begin{equation}
\frac{d\bar{B}(E1; \omega )}{d\omega }= \int  \! \! \! \! \! \! \! \sum _n
     B(E1; \omega _n)\rho (\omega -\omega _n )d\omega _n
\label{eq:be1_ave}
\end{equation}
where
\begin{equation}
\rho (\omega -\omega _n  )=\frac{1}{\pi }\frac{\Gamma /2}{(\omega
  -\omega _n )^2 +( \Gamma /2)^2 }.
\label{eq:weight}
\end{equation}
The weight factor can be considered to simulate the escape and the 
spreading widths. 
The width parameter  $\Gamma $ is arbitrary taken as 1MeV to draw a
smooth curve of the transition strength.
The oscillator length of the harmonic oscillator wave function is
taken as $b=(\hbar /m\omega _o )^{1/2}$=1.64 fm.  It is known that
the  photoreaction
cross section  $\sigma $ is related with the transition strength
$\bar{B}(E1; \omega ) $. The total photoreaction cross section
$\sigma _{int}$ 
%and the  first inverse energy moment $\sigma _{-1}$ 
is  written as\cite{BM75c} 
\begin{eqnarray}
\sigma _{int}=\int \sigma   d\omega = 
   \frac{16\pi ^3 }{9\hbar c}\int _0
^{E_{max}}\omega \frac{d\bar{B}(E1; \omega )}{d\omega }d\omega  
% \\
%\sigma _{-1}=\int \sigma  \omega ^{-1}d\omega =
% \frac{16\pi ^3 }{9\hbar c}\int _0
%^{E_{max}}\frac{d\bar{B}(E1; \omega )}{d\omega }  d\omega
\label{eq:photo}
\end{eqnarray}

The sum rule is a useful measure of the collectivity in GR.
For the IV GDR, the energy weighted sum rule value is given by 
\begin{equation}
S(TRK)= \int  \! \! \! \! \! \! \! \sum _n \hbar \omega _n 
  \mid \langle n|\hat{O}^{\lambda =1}_{\mu } |gs\rangle\mid ^2 
  =\frac{\hbar
  ^2}{2m} \frac{9}{4\pi } \frac{NZ}{A}    \nonumber \\
  = 14.9\frac{NZ}{A}e^2  \,\,\, (\mbox{MeV}\cdot\mbox{fm$^2$})
\label{eq:trk_sum}
\end{equation}
neglecting the contributions of exchange terms. 
This sum rule (\ref{eq:trk_sum}) 
is known as the classical Thomas-Reiche-Kuhn (TRK)
sum rule.  The cross section $\sigma _{int}$ is then expressed as
\begin{equation}
\sigma _{int}=\frac{16\pi ^3 }{9\hbar c}S(TRK)=60\frac{NZ}{A}
 \,\,\,\,(\mbox{MeVmb}).
\label{eq:trk_photo}
\end{equation}

The cluster sum rule is referred 
to measure the adiabaticity between GR and
Pigmy resonance\cite{SH90}.  
Assuming the valence cluster with $N_2$ and $Z_2$ 
and the core with $N_1$ and $Z_1$, the cluster sum rule is given 
 by
\begin{equation}
S(\mbox{cluster})
  =\frac{\hbar
  ^2}{2m} \frac{9}{4\pi } \frac{(Z_1 A_2 -Z_2 A_1)^2}{AA_1 A_2}  
\label{eq:cluster}
\end{equation}
where $A_1 =N_1 +Z_1$ and  $A_2 =N_2 +Z_2$.

%\section{Results}
The calculated results of averaged dipole strength (4) in C isotopes  are
shown in Figs. 8 and 10. The non-energy weighted summed transition strength
(NESR), the   energy weighted summed transition strength (EWSR) and 
the total photoreaction cross sections 
$\sigma _{int}$  
 are tabulated  in Table 3.  The low energy 
strength below GDR region is compared with  the cluster 
sum rule value (\ref{eq:cluster}) in Table 4.
The photoreaction cross sections for $^{12}$C, $^{13}$C and $^{14}$C
are also shown in Fig. 9. The width parameter is taken to be
$\Gamma$ = 2 MeV in Fig. 9.

In Fig. 8a, two calculated results of the transition strength in 
$^{12}$C obtained within 1$\hbar \omega $ and (1+3)$\hbar \omega $ 
configuration space are shown.  Although the peak energy of GDR appears 
at the same  energy E$_x \sim$ 21MeV, the integrated strength 
of the large space is 20\% smaller than that of the small space 
because of the coupling to many-particle many-hole states. The
experimental photoreaction cross section shows the GDR peak at 
E$_{x}$ = 22 MeV\cite{Pyw85} that is close to the present calculation.
The observed
$\sigma_{int}$ value exhausts 64\% of the TRK sum rule up to
E$_{x}$ = 30 MeV\cite{Pyw85}, while the
calculated sums show the  enhancement factor $\kappa $ for  the sum 
rule,i.e., 
 $\kappa $=0.62 (0.29) in the  1$\hbar \omega (3\hbar \omega )$ calculations.  
A large fraction of the strength is found to be in high excitation energy 
region up to E$_{x}$ = 140 MeV; $\kappa$ = 0.62 up to 100 MeV and $\kappa$
= 0.86 up to 140 MeV\cite{Ahr}. 
In ref. \cite{Ahr}, integrated cross section up to 35 MeV is found to be
about 65$\%\sim 90\%$ of the TRK sum rule, while calculated 
enhancement factor up to E$_{x}$ = 35 MeV is $\kappa$ = 0.01 in the (1+3)
$\hbar\omega$ space. In view of this, 25$\pm10\%$ 
of our calculated strength obtained within the (1+3) $\hbar\omega$ space 
should be in higher energy region than E$_{x}$
= 35 MeV. We, thus, need to 
reduce the calculated cross section for $^{12}$C by
multiplying a factor 0.7 as shown in Fig. 9 to obtain 
reasonable agreement with the available experimental one
in  ref. \cite{Pyw85}. 
The experiment data  show
the existence of 
a large fraction of the strength  in higher energy region
than $\bar{E_x }$=35MeV and suggest the importance of the coupling to
many-particle many-hole states more than 3$\hbar \omega $ excitations.
It  is interesting to notice that the difference between the two
calculations in Fig. 8a is only 200 keV for  the peak energy,  
 although the total cross section 
of 3$\hbar \omega $ calculation is  20\% smaller than that
of 1$\hbar \omega $ calculation.  

The calculated transition strength  $d\bar{B}(E1; \omega )/d\omega $ 
and the photoreaction cross section $\sigma $ in $^{13}$C ($^{14}$C)
obtained by including up to 3$\hbar\omega$ excitations are shown 
in Figs. 8b and 9b (Figs. 8c and 9c),  respectively. 
We  see  appreciable 
cross sections in both $^{13}$C and $^{14}$C below 14MeV, 
while there is essentially no cross section  in
$^{12}$C in the same energy region.  
The energy weighted sum (EWS) of the strength up to 14 MeV amounts to be 
86$\%$ and 66$\%$ of the cluster sum rule in $^{13}$C and $^{14}$C,
respectively, as shown in Table 4. The GDR peak appears at 25$\sim$26
MeV in $^{13}$C which is close to the experimental value at 24 MeV
\cite{Pyw85}. The distribution of the observed photoreaction cross
section is well reproduced by the present calculation within (1+3)
$\hbar\omega$ space. The integrated cross section calculated up to E$_{x}$ 
= 30 MeV
amounts to be 98$\%$ of the TRK sum rule, while the observed one is 
71$\%$ of the TRK value\cite{Pyw85}. About 30$\%$ of the calculated 
strength is
in the higher energy region beyond E$_{x}$ = 30 MeV, which is similar
to the case of $^{12}$C.

In case of $^{14}$C, the GDR peak appears at E$_{x}$ = 28 MeV which is
rather close to the observed one at 25.6 MeV\cite{Pyw85}. 
Experimental mean energies are 18.3$\pm$0.4 MeV and 26.7$\pm$0.1 MeV 
for T$_{<}$=1 and T$_{>}$=2 states, respectively\cite{Pyw85}, which are
close to our calculated values; 19.8 MeV for T$_{<}$=1 and 28.2 MeV
for T$_{>}$=2 states. The distribution of the photoreaction cross
section is well reproduced by the present calculation. 
Observed photoreaction cross sections for T$_{<}$ and T$_{>}$ states 
summed up to 30 MeV are 88$\pm$12 MeV$\cdot$mb  and 37$\pm$8 
MeV$\cdot$mb, respectively\cite{Pyw85}, while the present calculation
gives 119 MeV$\cdot$mb for T$_{<}$ states and 94 MeV$\cdot$mb for
T$_{>}$ states. Experimental values are quenched compared with the 
calculated ones by factors 0.74$\pm$0.1 for T$_{<}$ and 0.40$\pm$0.08
for T$_{>}$ states. A factor 0.4 is multiplied for T$_{>}$=2 states 
in Fig. 9. About 40$\pm10\%$ of the strength is found to be in the 
high energy region beyond E$_{x}$ = 30 MeV.  
  
Calculated dipole strength for $^{15}$C, $^{16}$C, $^{17}$C, 
$^{18}$C and $^{19}$C are shown in Fig. 10.
Since the spin and the parity of  the ground state in 
$^{19}$C is not established yet 
experimentally,  we calculate the dipole strength excited from 
two possible
spin-parity states 1/2$^{+}$ and  3/2$^{+}$ for the ground state.   
In case of $^{15}$C, effects of skin are studied. The neutron
1s$_{1/2}$-orbit is obtained in a Woods-Saxon well to reproduce
the experimental separation energy of 1.22 MeV. The dipole 
strength enhanced about by 30$\%$ in the low energy region around
E$_{x}$ = 5 MeV. The skin effect is rather moderate.
The NESW and EWSR of the 
transition 
strength B(E1) and the total photoreaction cross sections are 
listed in Tables 3 and 4. The dipole strength below GDR region 
becomes substantial in these nuclei, i.e., the cross sections
$\sigma_{int}$ below $\hbar\omega$=14 MeV exhaust 7.8$\%$ for $^{15}$C ,
 16.3$\%$  for $^{16}$C ,
13.1$\%$   for $^{17}$C, 11.6$\%$  for $^{18}$C and 12.6$\%$ for 
1/2$_{g.s.}^{+}$ case of  $^{19}$C 
( 14.1 $\%$ for 3/2$_{g.s.}^{+}$ case of $^{19}$C) of the TRK sum rule, 
respectively. These values correspond to 46.6$\%$ for $^{15}$C ,
 81.3$\%$  for $^{16}$C ,
57.8$\%$   for $^{17}$C, 46.4$\%$  for $^{18}$C and 47.0$\%$ for 
1/2$_{g.s.}^{+}$ case of  $^{19}$C 
( 52.4  $\%$ for 3/2$_{g.s.}^{+}$ case of $^{19}$C) of the cluster sum rule, 
respectively.
The GDR peaks with the isospin  $T_{<}$ are found
always at around  E$_{x}$= 17$\sim$19 MeV in these nuclei. 
On the other hand,  the  $T_{>}$ peaks appear  more than 10 MeV
higher in energy than the  $T_{<}$ peaks,  
and smaller in peak height in heavier
C-isotopes. In the extreme case of $^{19}$C, 
the cross section $\sigma_{int}$ of
$T_{>}$ states becomes only 13.5\% of the TRK sum rule and there are only
very small strength of  $T_{>}$ states below  E$_{x}$=30MeV.

There are peaks at rather low energies 10 - 20 MeV in $^{15\sim19}$C. 
These energies are close to the unperturbed p$-$h excitation energy
  1$\hbar \omega$(H.O.)=41/A$^{1/3}$ MeV in the harmonic oscillator model, 
but  much  lower than the  systematic excitation energies of 
 Giant resonances $\hbar \omega $(GDR)$\sim $80/A$^{1/3}$ MeV.  
This low energy feature might be  attributed to the effects of large 
deformations (see Table 1), 
  which makes some  unperturbed p-h 1$^-$ states lower than those of spherical 
nuclei. The main GR part have two peaks in $^{15}$C and $^{16}$C. This can
 be also considered as the effects of the strong prolate deformation as 
is seen in Table 1.
 In cases of
$^{17}$C,  $^{18}$C and $^{19}$C, 
 any clear two-peak structure is not seen in the strength distributions 
while 
%somesymptom looks to remain in $^{18}$C.
the main peaks have large widths of  $\Gamma \sim $10MeV.
The strength  distributions show 
 not much difference between the case of spin 1/2$^{+}$ 
for the ground state (prolate deformation) and that of spin 3/2$^{+}$
for the ground state (oblate deformation).  
As the strength around the peak up to 12-14 MeV exhausts about 50\% of the 
cluster sum rule value in the heavier C isotopes, 
these regions may be interpreted as pigmy resonances.
The strength is fragmented widely in the heavier isotopes , 
 and the distinction between Giant resonance and
pigmy resonance seems not very clear except for $^{15}$C and $^{18}$C.

\section{SUMMARY}
 We have studied the ground state properties of C
isotopes by the deformed Hartree-Fock +BCS model. The shallow
deformation minima are found in the several isotopes. The prolate
deformation is suggested  to favor for $^{15}$C and $^{17}$C while the 
oblate deformation is most probable 
 for $^{19}$C. Both the prolate and the 
oblate minima appear in $^{16}$C and  $^{18}$C to be 
almost degenarate in energy.

The magnetic and the Q- moments of odd isotopes are investigated
by the shell model calculations, and their configuration dependence
is pointed out. It is crucial  to obtain  experimental information
on the values of magnetic and quadrupole moments to determine
whether the deformation is prolate or oblate. In particular, 
it would be interesting to find out  decisive information on the
deformation of $^{19}$C since  this nucleus is a keystone 
to establish  the new shell
structure at N=16  in the C isotopes.  

We have also studied the Pigmy and GDR dipole strengths of
C-isotopes by using shell model calculations in the  large
scale shell model (0p-1s0d-1p0f) space. We found that the excitation
energies of  GDR in $^{12}$C,  $^{13}$C and $^{14}$C show  good agreement
with the experimental data of the  two isospin resonances,  T$_>$
and T$_<$.
Moreover the calculated Pigmy strength below $\hbar
\omega =$14 MeV in $^{14}$C  is consistent with the experimental photoreaction cross
sections.
In heavier  C-isotopes than  $^{14}$C,   the T$_< $ GDR
has always a peak at around E$_{x}$ = 17$\sim$19 MeV, 
while the T$_> $ peak is more than 10 MeV higher in 
energy and much smaller in the cross section than the T$_< $ one.
In these heavy isotopes, the
Pigmy resonances are more pronounced  than that of  $^{14}$C,
having  about 8$\sim16\%$ of the TRK sum rule  values, which correspond
to  50\% of the cluster sum rule values .
Future experimental effort is highly desirable to observe these
Pigmy resonances to clarify the structure of drip line nuclei\cite{Emi98}.

\section*{Acknowledgements}
 We thank P.-G. Reinhard for providing us with his computer code 
for the deformed Hartree-Fock + BCS calculations and also for his 
constant help in running the code.
This work was supported in part  by the Japanese Ministry of Education,
Science, Sports and Culture by Grant-In-Aid for Scientific Research 
 under the
program number C(2) 12640284.

\newpage
%{\bf References}
%\section*{References}

\newpage
{\bf Table 1}\,\,\,
The energy minima of the energy surface 
in the deformed HF calculations with the Skyrme
interaction SkI4: (a) with the volume-type delta pairing interaction in Eq.  
(\ref{eq:pair_v}), and (b)  with the surface-type 
density-dependent  pairing interaction in Eq.  
 (\ref{eq:pair_s}). 
\begin{center}
  \begin{tabular}{ c|c|c|c|c||c|c|c|c } \hline
 (a)  & & & & & (b) & & & \\ \hline
 nucleus &  K$^{\pi}$ &   $\beta_2 $ & Energy & ${\bar{\Delta} _n }$ &
K$^{\pi}$ &   $\beta_2 $ & Energy & ${\bar{\Delta} _n }$ \\
 &   &   & (MeV) & (MeV)
 &   &   & (MeV) & (MeV) \\
\hline \hline
 $^{12}$C & 0$^{+}$ & 0.0   &  $-$88.54  &  0.0 & 
0$^+$ & 0.0 & $-88.54$ & 0.0 \\ 
\hline
 $^{13}$C & $\frac{1}{2}^{-}$ & 0.0   &  $-$97.12  &  0.0 
& $\frac{1}{2}^{-}$ & 0.0   &  $-$97.12  &  0.0 \\ 
       & $\frac{1}{2}^{+}$ &$-$0.34    & $-$90.18  &  0.0 
       & $\frac{1}{2}^{+}$ &$-$0.34    & $-$90.18  &  0.0 \\ 
   & $\frac{1}{2}^{+}$ & 0.33   &  $-$92.27  &  0.0 
   & $\frac{1}{2}^{+}$ & 0.30   &  $-$91.86  &  1.06 \\ 
   & $\frac{5}{2}^{+}$ & $-$0.34   &  $-$91.03 &  0.0 
   & $\frac{5}{2}^{+}$ & $-$0.34   &  $-$91.03 &  0.0 \\ \hline
 $^{14}$C & 0$^{+}$ & 0.0  &$-$106.7 &  0.0 
& 0$^{+}$ & 0.0  &$-$106.7 &  0.0 \\\hline
  $^{15}$C & $\frac{1}{2}^{+}$ &$-$0.084    & $-$107.8   &  0.0 
& $\frac{1}{2}^{+}$ &$-$0.084    & $-$107.8   &  0.0 \\
      & $\frac{1}{2}^{+}$ &$+$0.195    & $-$109.7   &  0.0 
      & $\frac{1}{2}^{+}$ &$+$0.195    & $-$109.7   &  0.0 \\
  &  $\frac{5}{2}^{+}$   & $-$0.123  & $-$108.2  &  0.0   
  &  $\frac{5}{2}^{+}$   & $-$0.123  & $-$108.2  &  0.0   \\\hline
   $^{16}$C &   0$^{+}$&    $-$0.176 &    $-$109.8 &    1.08   
 &   0$^{+}$&    $-$0.10 &    $-$111.72 &    2.36   \\
             &   0$^{+}$   &   0.298   &   $-$110.5   & 0.705 
             &   0$^{+}$   &   0.108   &   $-$112.0   & 2.30 \\   \hline
  $^{17}$C &  $\frac{1}{2}^{+}$ &   $-$0.224  &   $-$111.5 &       1.04
&  $\frac{1}{2}^{+}$ &   $-$0.184  &   $-$112.3 &    2.14  \\
         &$\frac{1}{2}^{+}$  &  0.241  &    $-$110.9  &   0.848  
         &$\frac{1}{2}^{+}$  &  0.138  &    $-$111.9  &   2.105  \\
 & $\frac{3}{2}^{+}$  &  $-$0.188 &    $-$110.7 &       1.02  
 & $\frac{3}{2}^{+}$  &  $-$0.184 &    $-$111.4 &       2.14  \\
 &  $\frac{3}{2}^{+}$    &  0.375 &    $-$112.2 &       0.0   
 &  $\frac{3}{2}^{+}$    &  0.333 &    $-$111.95 &       1.67   \\
   & $\frac{5}{2}^{+}$  &  $-$0.224 &   $-$111.3 &       1.04  
   & $\frac{5}{2}^{+}$  &  $-$0.221 &   $-$111.6 &       1.20  \\   \hline
 $^{18}$C & 0$^{+}$   &  $-$0.273  &   $-$113.4 &  0.930  
 & 0$^{+}$   &  $-$0.238  &   $-$115.0 &  2.14  \\
 &  0$^{+}$   & 0.345  &    $-$113.8   &  0.0    
 &  0$^{+}$   & 0.191  &    $-$115.3   &  2.16    \\   \hline
 $^{19}$C & $\frac{1}{2}^{+}$   &  0.326   &  $-$114.3    &    0.246
& $\frac{1}{2}^{+}$   &  0.155   &  $-$114.8    &    2.05   \\
           &  $\frac{3}{2}^{+}$   &  $-$0.301 &    $-$115.0  &      0.914
           &  $\frac{3}{2}^{+}$   &  $-$0.293 &    $-$114.8  &      1.91 \\   \hline
 $^{20}$C & 0$^{+}$   &  $-$0.299  &   $-$117.3 &  0.0  
 & 0$^{+}$   &  $-$0.232  &   $-$117.3 &  1.06  \\ \hline
$^{22}$C & 0$^{+}$   &  0.00  &   $-$118.2 &  0.0       
 & 0$^{+}$   &  0.00  &   $-$118.19 &  1.12  \\ \hline

\end{tabular}
\end{center}

\newpage
%\vspace*{3cm}
\noindent
{\bf Table 2}\,\,\,
Magnetic moments and quadrupole moments of C-isotopes.
Shell model calculations are performed
with the Warburton-Brown WBP10 interaction. The effective spin g-factor for neutron 
is taken to be $g_s (eff)$/$g_s (bare) $=0.9 in the shell model calculations.
The effective charges for Q-moments are taken from the results of 
particle-vibration 
model based on HF+RPA calculations in ref. \cite{SA01}.  The 
experimental data of $g-$factors are taken from ref. \cite{C15mm} for $^{15}$C
and from ref. \cite{C17mm} for  $^{17}$C.  The single particle Q-moment is
calculated by using the harmonic oscillator wave function with the
oscillator length b=1.64 $fm$.
\begin{center}
  \begin{tabular}{ c|c|c|c|c|c|c|c } \hline
A & J$^{\pi}$   & Energy  & g(Schmidt) &g(cal) & $|$g(exp) $|$  
& Q-moment(s.p.) & Q-moment(cal) 
   \\
  &  & (MeV) & $ (\mu _N )$ & $ (\mu _N )$   &$ (\mu _N )$  & $e $ mb 
& $e $ mb  \\\hline
 $^{15}$C &  $\frac{1}{2}^{+}$ & 0.00 & -3.83 & -3.37 &   & $----$  &  \\\hline
    &      &       &         &     & 3.440  $\pm$ 0.018 &  &    \\\hline
 $^{17}$C &  $\frac{1}{2}^{+}$ & 0.295 & -3.83 & -2.82 &   & $----$  &  \\
         &  $\frac{3}{2}^{+}$ & 0.00 & 0.765 & -0.514 &   & -37.7$e_{eff}(n)$ & 23.5    \\
 &  $\frac{5}{2}^{+}$ & 0.032 &  -0.765 &-0.505 &    & -53.8$e_{eff}(n)$  &   \,\,-9.3  \\\hline
 &       &   &    &   & 0.5054 $\pm$0.0025  &  \\\hline
$^{19}$C &  $\frac{1}{2}^{+}$ & 0.00 & -3.83 & -2.600  &   &  \\
 &  $\frac{3}{2}^{+}$ & 0.625 &  0.765 & 0.187  &    &  -37.7$e_{eff}(n)$   &  -33.1     \\
 &  $\frac{5}{2}^{+}$ & 0.190 & -0.765 & -0.411  &    &  -53.8$e_{eff}(n)$ & \,\,1.1   \\\hline
\end{tabular}
\end{center}

\newpage
\noindent
{\bf Table 3}\,\,\,
 Non-energy weighted sum rule (NESR) and energy weighted sum rule 
(EWSR) values
of E1 transitions in C-isotopes. 
Integrated photoreaction cross sections 
$\sigma _{int}$ (MeV$\cdot$mb) are also shown in the Table. 
Shell model calculations are performed
with the Warburton-Brown WBP10 interaction. 
\begin{center}
  \begin{tabular}{ c|c|c|c|c|c|c|c } \hline
 % \begin{tabular}   \hline
%\multicolumn{1}{c} {A} & \multicolumn{1}{c} {Isospin}  &  \multicolumn{1}{c}
% {NEWS (fm$^{2} $)} &  \multicolumn{1}{c} {EWSR (MeV$\cdot$fm$^{2} $)} &
% \multicolumn{1}{c} {$\bar{E_{x} }$} &  \multicolumn{1}{c}
A & Isospin & NEWS & EWSR  & $\bar{E_{x} }$ &S(TRK) & EWSR/S(TRK) & $\sigma_{int}$ \\\hline
  &         &    $e^2 $fm$^{2} $& MeV$\cdot e^2 $fm$^{2} $ & MeV
 & MeV$\cdot e^2 $fm$^{2} $ & \%   & MeV$\cdot $ mb  \\\hline
$^{12}$C ($1\hbar \omega$)& T$ =1$ & 2.89 & 72.3  & 25.0 &44.7 & 161.7
   & 291. \\\cline{2-8}
 ($(1+3)\hbar \omega$)& T$ =1$&2.29&57.6&25.2 &44.7 &
  128.8&232. \\\hline
 $^{13}$C& T$_< =1/2$&1.29&25.7&19.9&  & &103. \\
 ($1\hbar\omega$)  & T$_> =3/2$ & 1.81 & 46.7 & 25.8  &  & & 188. \\\cline{2-8}
    & total &   3.10   & 72.4 &       & 48.1  & 151. & 291.  \\\hline
 $^{13}$C&  T$_< =1/2$&1.12&22.4&20.0&  & &  90.2 \\
 ($(1+3)\hbar\omega$)& T$_> =3/2$ & 1.39 & 35.6 & 25.6 & & & 144. \\\cline{2-8}
    & total & 2.51 & 58.0 &  &  48.1 & 121. & 234. \\\hline
 $^{14}$C & T$_< =1$ & 2.02 & 41.1 & 20.4&  &  & 166. \\
 ($(1\hbar\omega$) & T$_> =2$ & 1.28 & 35.7 & 27.8 & & & 144. \\\cline{2-8}
   & total & 3.30 & 76.8 &  & 51.1 & 150. & 309. \\\hline 
 $^{14}$C& T$_< =1$&1.71&33.7&19.7&  & &136. \\
 ($(1+3)\hbar \omega$)    & T$_> =2$ & 0.958 & 27.0 & 28.2 &  & & 109. \\\cline{2-8}
    & total &   2.67  & 60.7 &       & 51.1 & 119. & 245.  \\\hline
 $^{15}$C &  T$_< =3/2$ & 2.63 & 48.9 & 18.6 &  & & 197.   \\
   &  T$_> =5/2$        &  0.866& 25.1  & 29.0 & & & 101.  \\\cline{2-8}
    & total &       3.49 & 74.1 &   &53.6 & 138. & 298. \\\hline
 $^{16}$C &   T$_< =2$ & 2.77  & 50.3 & 18.2 & && 203.  \\
    &   T$_> =3$ &  0.651 & 20.3 &  31.1& & & 81.7  \\\cline{2-8}
   & total &        3.42 & 70.6 &  & 55.9 & 126. & 284. \\\hline
 $^{17}$C &   T$_< =5/2$ & 3.14  & 57.1 & 18.2 & && 230.  \\
    &   T$_> =7/2$ &  0.460 & 14.5 &  31.6& & & 58.3  \\\cline{2-8}
   & total &        3.60 & 71.7 &  & 57.9 & 124. & 288. \\\hline
 $^{18}$C &   T$_< =3$ & 3.26 & 59.6 & 17.5 & && 240. \\ 
     &   T$_> =4$ & 0.327&  11.3 & 34.6 &  && 45.5 \\ \cline{2-8}
     & total       & 3.59 & 68.5 & & 59.6 & 115. & 276. \\   \hline
 $^{19}$C & T$_< =7/2$ & 3.46 & 60.1 & 17.4 & & & 242. \\
 1/2$_{g.s.}^{+}$ &  T$_> =9/2$  & 0.23 & 8.2 & 35.4 & & & 33.0 \\\cline{2-8}
     & total       & 3.69 & 68.3 & & 61.2 & 112. & 275. \\\hline 
 $^{19}$C & T$_< =7/2$ & 3.42 & 57.7 & 16.9 & & & 232. \\
 3/2$_{g.s.}^{+}$  & T$_> =9/2$  & 0.22 & 7.8 & 35.0 & & & 31.3 \\\cline{2-8}
     & total       & 3.64 & 65.5 & & 61.2 & 107. & 264. \\\hline 
\end{tabular}
\end{center}

\newpage
{\bf Table 4}\,\,\,
Low energy strength of electric  dipole 
transitions in C isotopes.
Energy weighted sum rule (EWSR) values are compared with 
the cluster sum rule values $\sigma _{clu}$.
Shell model calculations are performed
with the Warburton-Brown WBP10 interaction. 
\begin{center}
  \begin{tabular}{ c|c|c|c|c|c|c } \hline
 A  &  $\sigma _{clu}$ &  \multicolumn{3}{|c|}{EWSR(MeV$\cdot$fm$^{2} $)}\\
\hline
  &   & E$_{x} <$ 12MeV (\%)   &  E$_{x} <$ 14MeV (\%) &
  E$_{x}<$ 16MeV (\%)\\\hline
$^{12}$C ($(1+3)\hbar \omega$)& 0.0 & 0.13 ($--$) & 0.23 ($--$) & 0.43($--$) \\\hline
 $^{13}$C ($(1+3)\hbar\omega$)  &3.44&   2.02 (58.7)  & 2.97 (86.3)    & 4.01 (116.6)    \\\hline
 $^{14}$C ($(1+3)\hbar \omega$)& 6.39 & 1.14 (17.8)  & 4.22 (66.0) & 9.24 (145.) \\\hline
  $^{15}$C &8.94     &2.28 (25.5) &4.17 (46.6)  &12.93 (144.6)   \\\hline
   $^{16}$C &  11.18  &  2.57 (23.0)  &  9.09 (81.3) &  15.46 (138.)  \\\hline
 $^{17}$C   & 13.15   & 3.56 (27.1)    & 7.60 (57.8)   & 13.48 (102.5)   \\\hline
  $^{18}$C &  14.90  &  3.08 (20.7) &  6.91 (46.4)  &  12.82 (86.0)  \\\hline
 $^{19}$C (1/2$_{g.s.}^{+}$)  & 16.47   & 4.00 (24.3)    & 7.74 (47.0)   
 & 13.54 (82.2)  \\\hline
 $^{19}$C (3/2$_{g.s.}^{+}$)  & 16.47   & 4.63 (28.1)   & 8.64 (52.4)  
 & 15.15 (92.0) \\\hline
 \end{tabular}
\end{center}

%\newpage
%\noindent
%{\bf Figure Captions}

\noindent
\newpage

\begin{figure}[tbh]
\begin{center}
%\psbox[scale=0.4]{fig2_gqr.ps}
%\epsf{file=fig2_gqr.ps,height=3.4in,width=3.4in}
%\psbox[scale=0.4]{fig2_gqr.ps}
\psfig{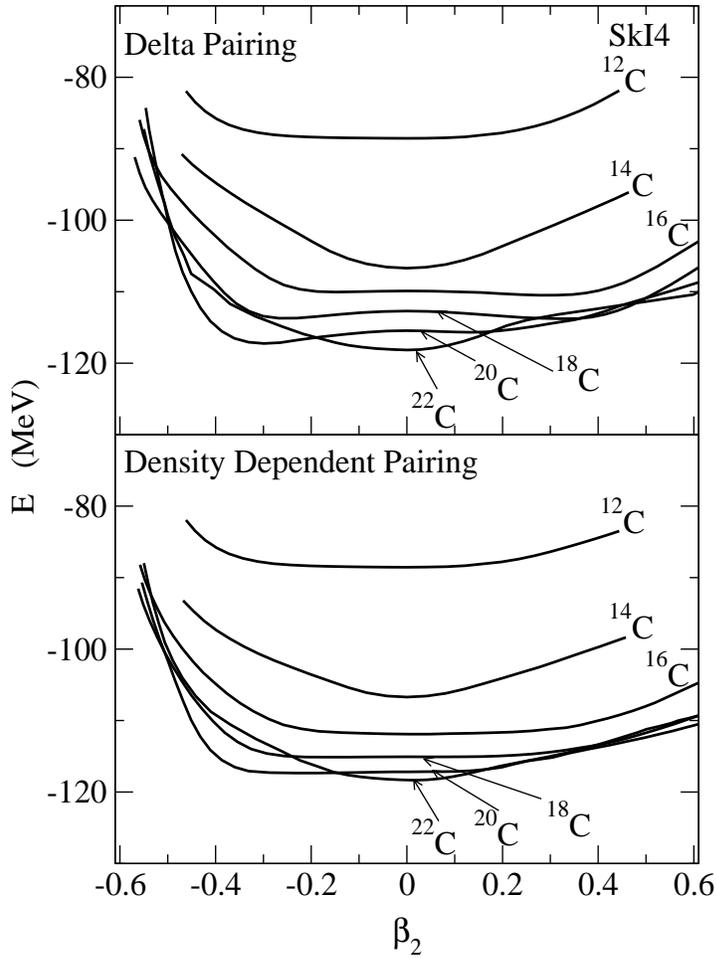}
\caption{ The energy surfaces for  the ground state
of even-mass C isotopes obtained by the 
HF$+$BCS calculations
with a Skyrme interaction SkI4 together with 
the volume-type delta pairing interaction (upper panel) in 
 Eq. (\ref{eq:pair_v})
 and the surface-type density-dependent pairing interaction 
(lower panel)  in Eq. (\ref{eq:pair_s}).
\label{fig:c12_def}}
\end{center}
\end{figure}

\begin{figure}[tbh]
\begin{center}
%\psbox[scale=0.4]{fig2_gqr.ps}
%\epsf{file=fig2_gqr.ps,height=3.4in,width=3.4in}
%\psbox[scale=0.4]{fig2_gqr.ps}
\psfig{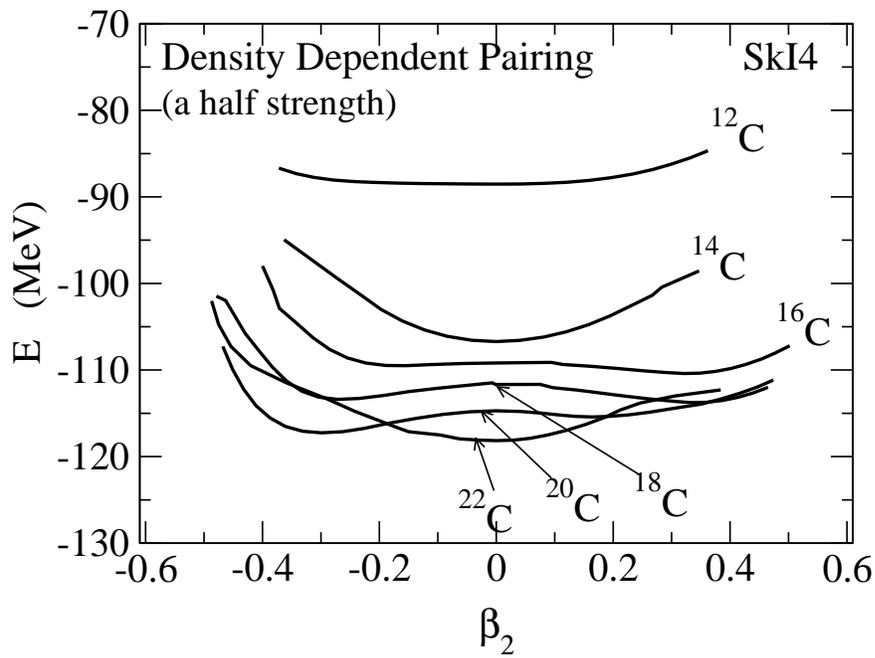}
\caption{ The energy surface for  the ground state
of  even-mass C isotopes obtained by the 
HF$+$BCS calculations
with a 
   Skyrme interaction SkI4 together with 
the surface-type pairing interaction   (\ref{eq:pair_s}) 
 with the weak pairing strength  
 $V_0^{'}=-500$ MeV$\cdot$fm$^3$ for neutrons and $-573
 $ MeV$\cdot$fm$^3$ for protons.  
\label{fig:c12_def2}}
\end{center}
\end{figure}

\begin{figure}[tbh]
\begin{center}
\hspace{0.5cm}
%\psbox[scale=0.4]{fig2_gqr.ps}
%\epsf{file=fig2_gqr.ps,height=3.4in,width=3.4in}
%\psbox[scale=0.4]{fig2_gqr.ps}
\psfig{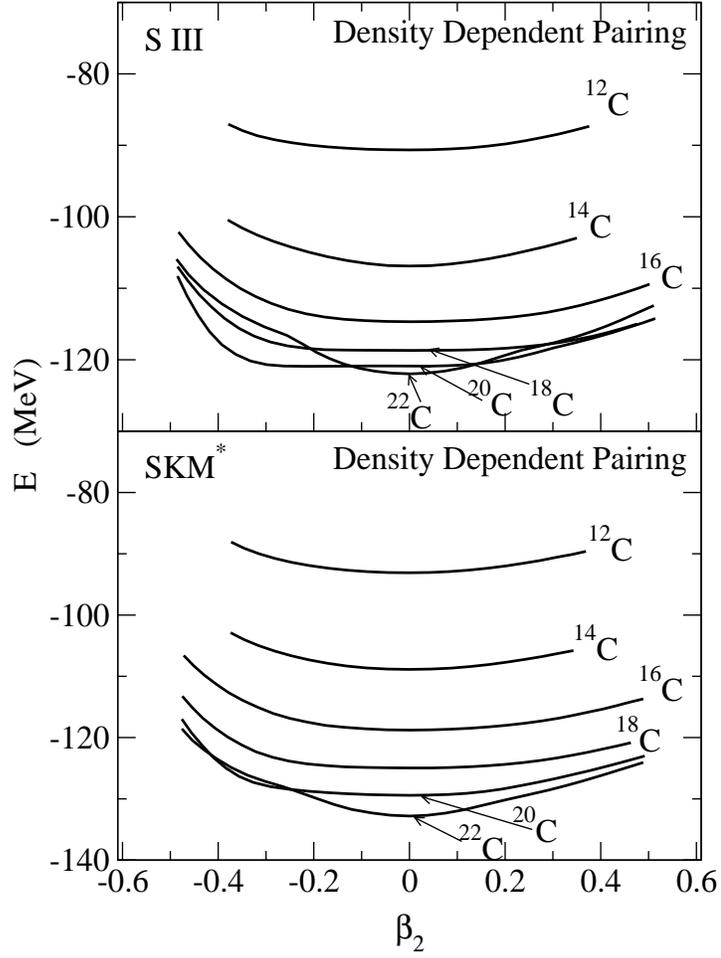}
\caption{The energy surfaces for  the ground state
of  even-mass C isotopes obtained by the 
HF$+$BCS calculations
with Skyrme interaction 
SIII (upper panel) and SkM$^*$ (lower panel). 
The surface-type  density dependent pairing interaction  (\ref{eq:pair_s})
   is used for the BCS calculations. } 
\end{center}
\end{figure}

\begin{figure}[tbh]
\begin{center}
\hspace{1cm}
%\psbox[scale=0.4]{fig2_gqr.ps}
%\epsf{file=fig2_gqr.ps,height=3.4in,width=3.4in}
%\psbox[scale=0.4]{fig2_gqr.ps}
\psfig{figure=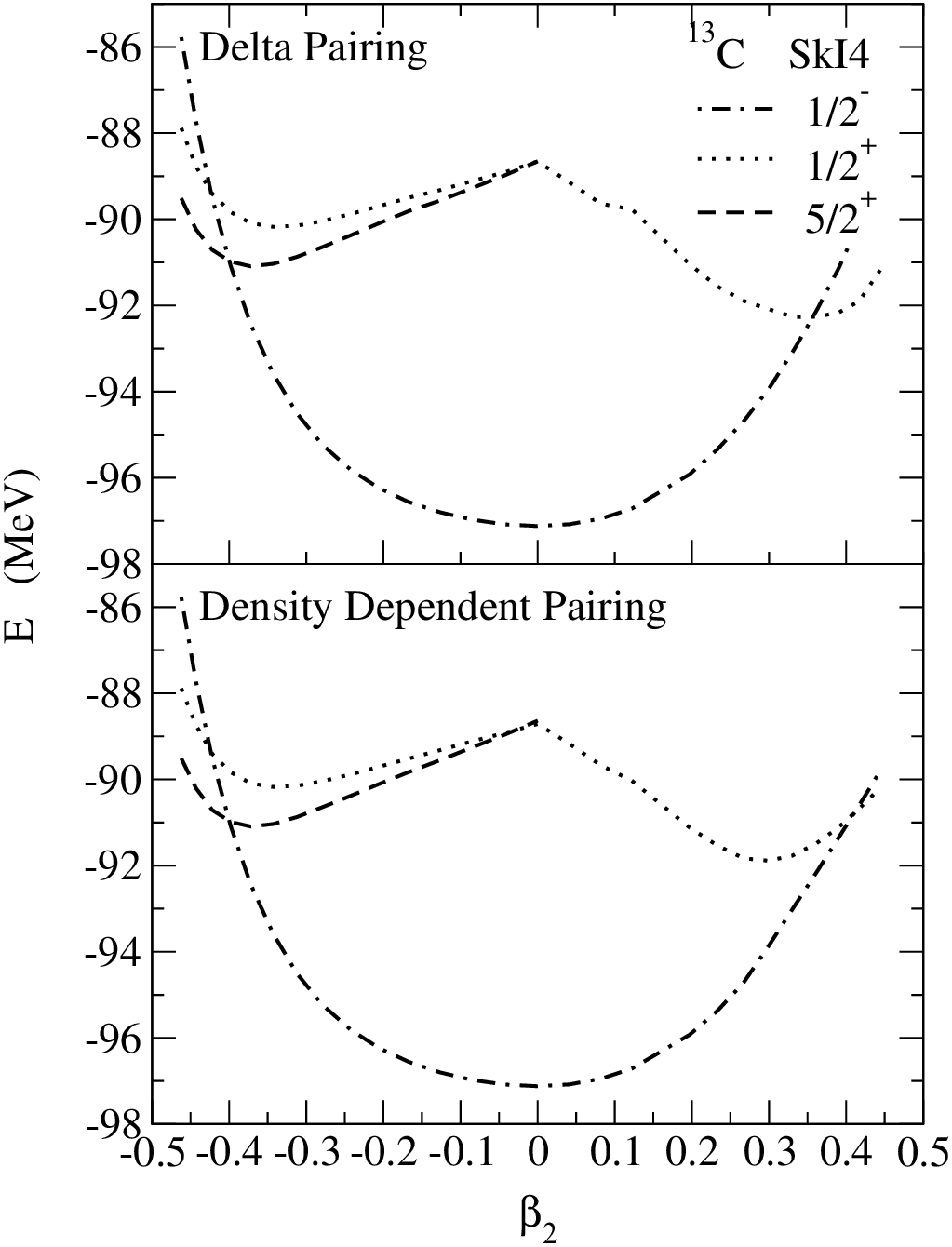,height=5.0in}
\caption{One quasi-particle energy surface for $^{13}$C  on top of the 
BCS ground state of  $^{12}$C. 
The SkI4 interaction with 
the volume-type delta pairing interaction (upper panel) and the surface-type 
density-dependent pairing interaction (lower panel)
   are used for the HF+BCS calculations. }
\end{center}
\end{figure}

\begin{figure}[tbh]
\begin{center}
\hspace{1cm}
%\psbox[scale=0.4]{fig2_gqr.ps}
%\epsf{file=fig2_gqr.ps,height=3.4in,width=3.4in}
%\psbox[scale=0.4]{fig2_gqr.ps}
\psfig{figure=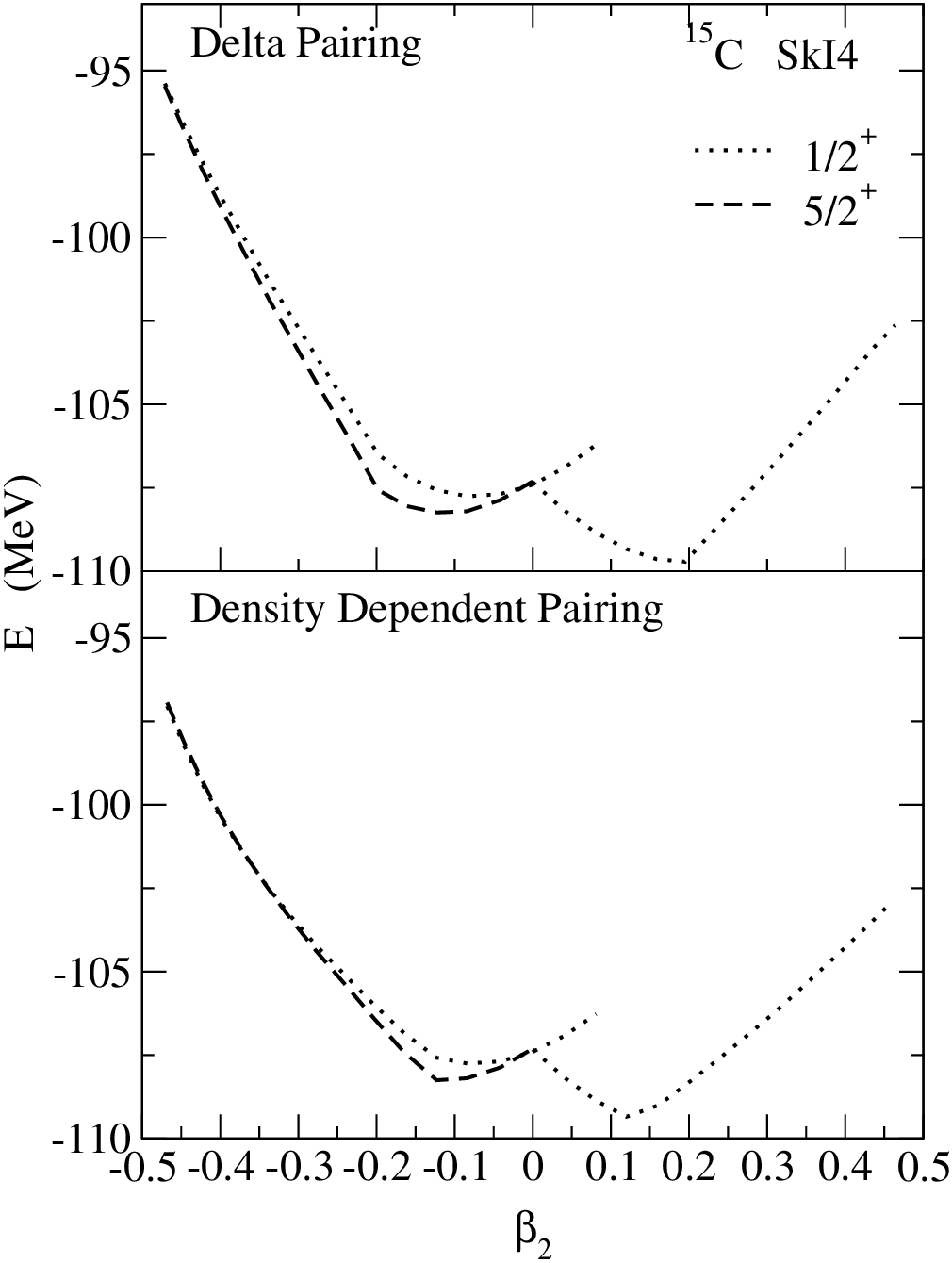,height=5.0in}
\caption{One quasi-particle energy surface for  $^{15}$C on top of the 
BCS ground state of  $^{14}$C. 
The SkI4 interaction with 
the volume-type delta pairing interaction (upper panel) and the surface-type 
density-dependent pairing interaction (lower panel)
   are used for the HF+BCS calculations.}
\end{center}
\end{figure}

\begin{figure}[tbh]
\begin{center}
\hspace{1cm}
%\psbox[scale=0.4]{fig2_gqr.ps}
%\epsf{file=fig2_gqr.ps,height=3.4in,width=3.4in}
%\psbox[scale=0.4]{fig2_gqr.ps}
\psfig{figure=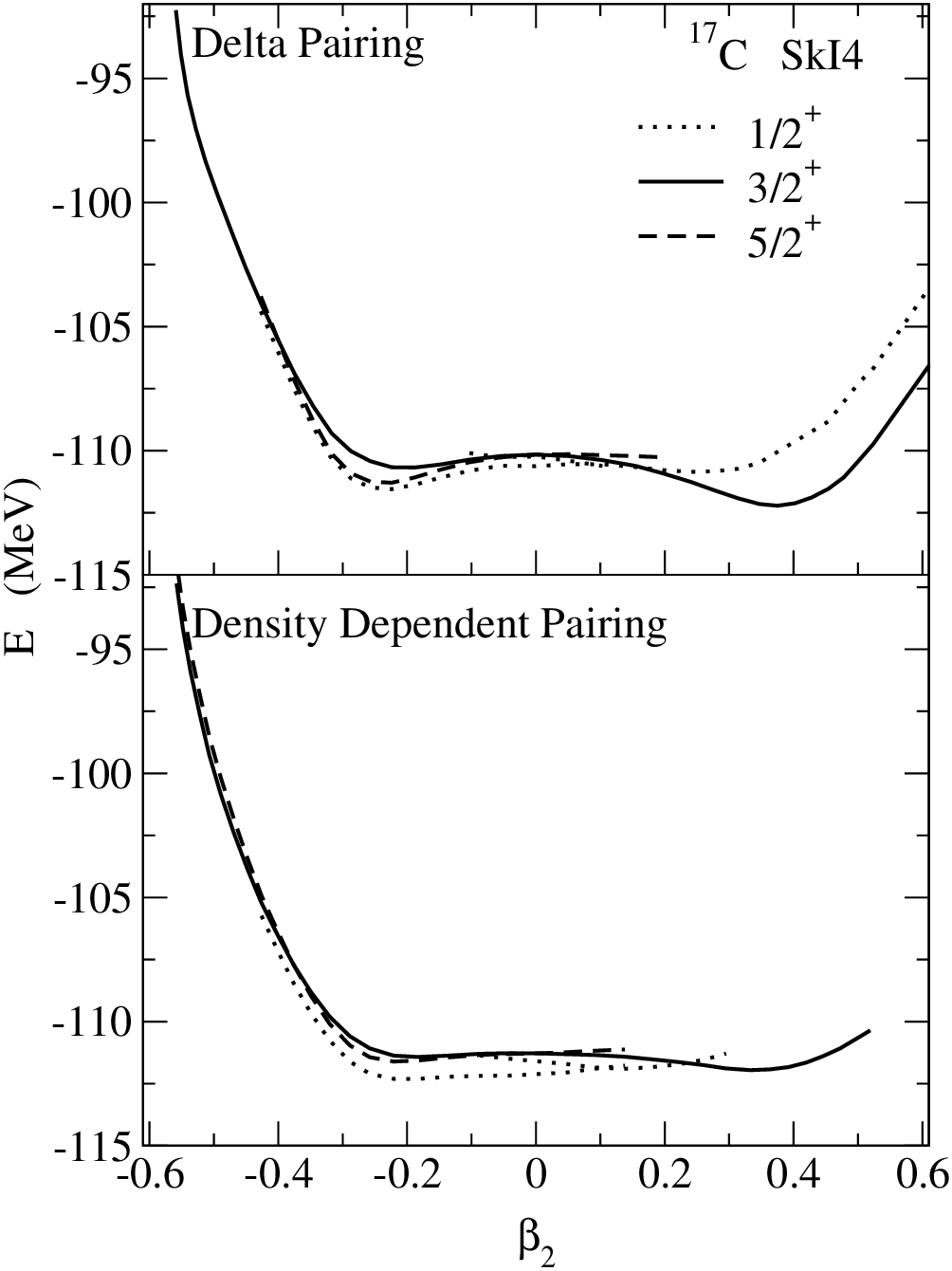,height=5.0in}
\caption{One quasi-particle energy surface for $^{17}$C on top of the 
BCS ground state of  $^{16}$C. 
The SkI4 interaction with 
the volume-type delta pairing interaction (upper panel) and the surface-type 
density-dependent pairing interaction (lower panel)
   are used for the HF+BCS calculations.}
\end{center}
\end{figure}

\begin{figure}[tbh]
\begin{center}
\hspace{1cm}
%\psbox[scale=0.4]{fig2_gqr.ps}
%\epsf{file=fig2_gqr.ps,height=3.4in,width=3.4in}
%\psbox[scale=0.4]{fig2_gqr.ps}
\psfig{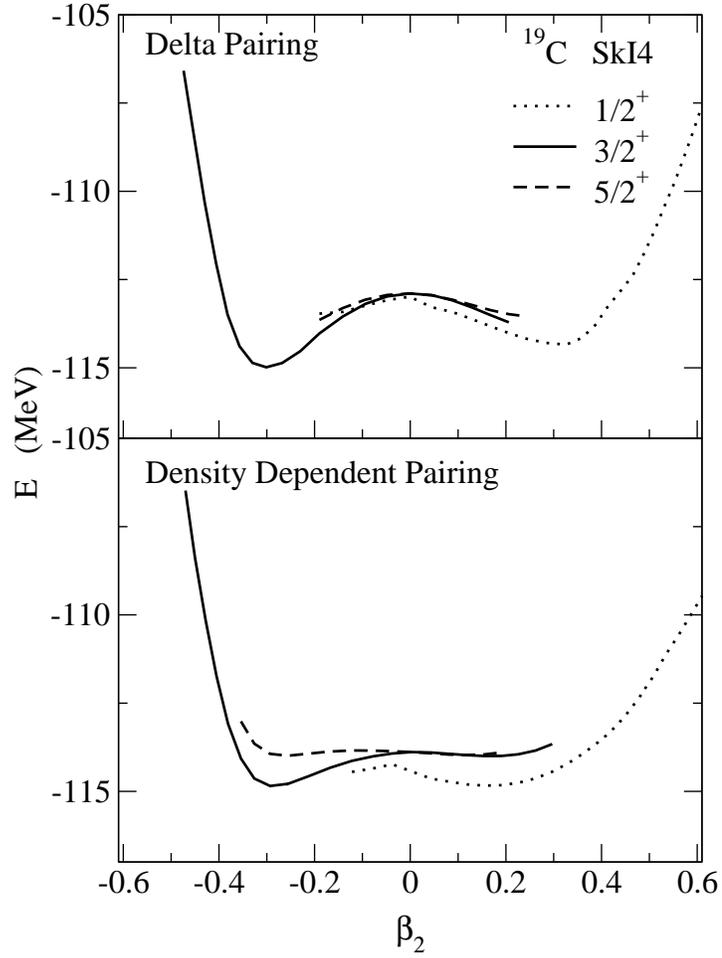}
\caption{One quasi-particle energy surface for $^{19}$C on top of the 
BCS ground state of $^{18}$C. 
The SkI4 interaction with 
the volume-type delta pairing interaction (upper panel) and the surface-type 
density-dependent pairing interaction (lower panel)
   are used for the HF+BCS calculations.}
\end{center}
\end{figure}

s
\begin{figure}[tbh]
\begin{center}
\vspace{-10mm}
\hspace{-20mm}
\psfig{figure=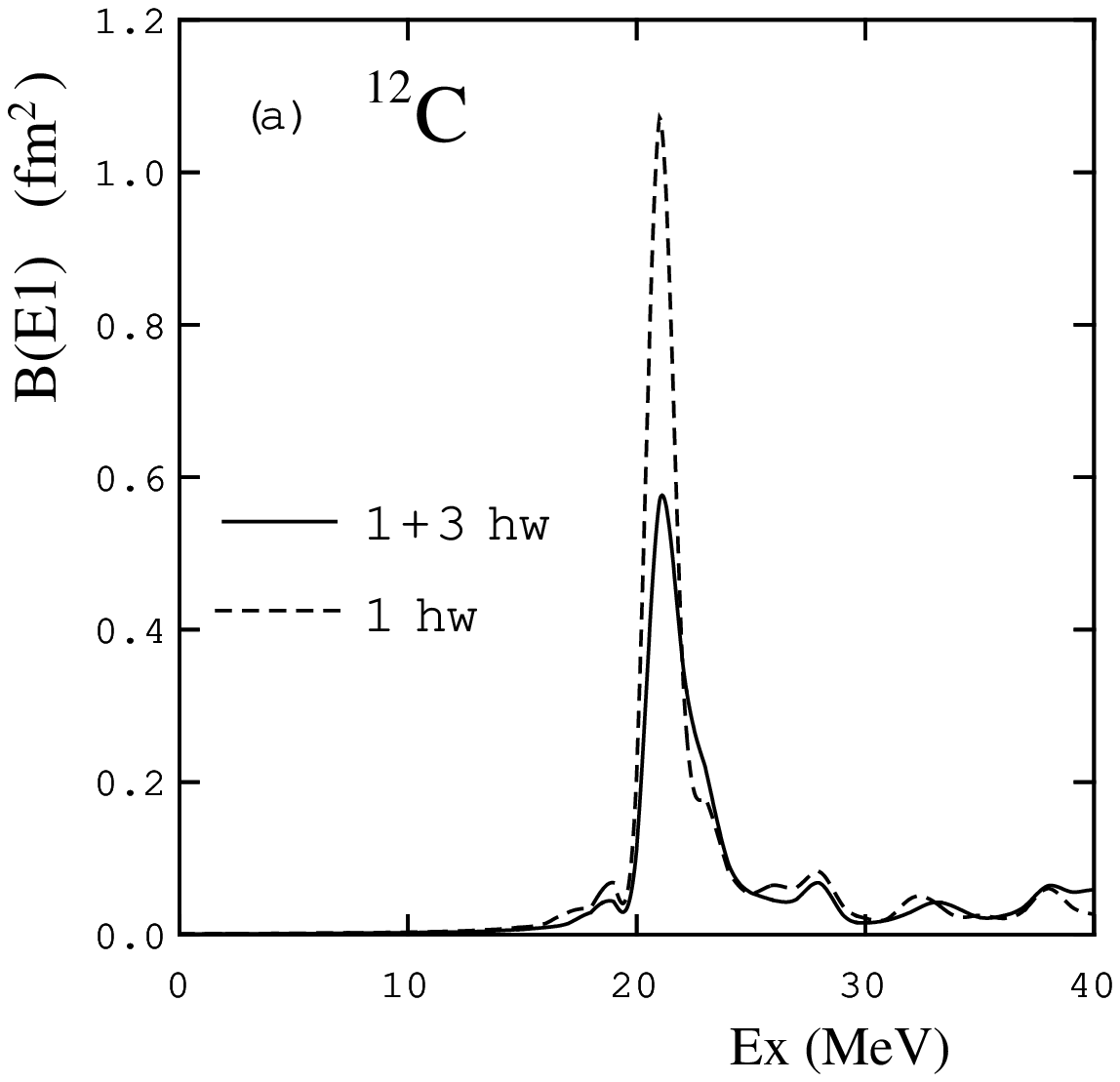,height=3.6in}
\vspace{-10mm}
\hspace{-20mm}
\psfig{figure=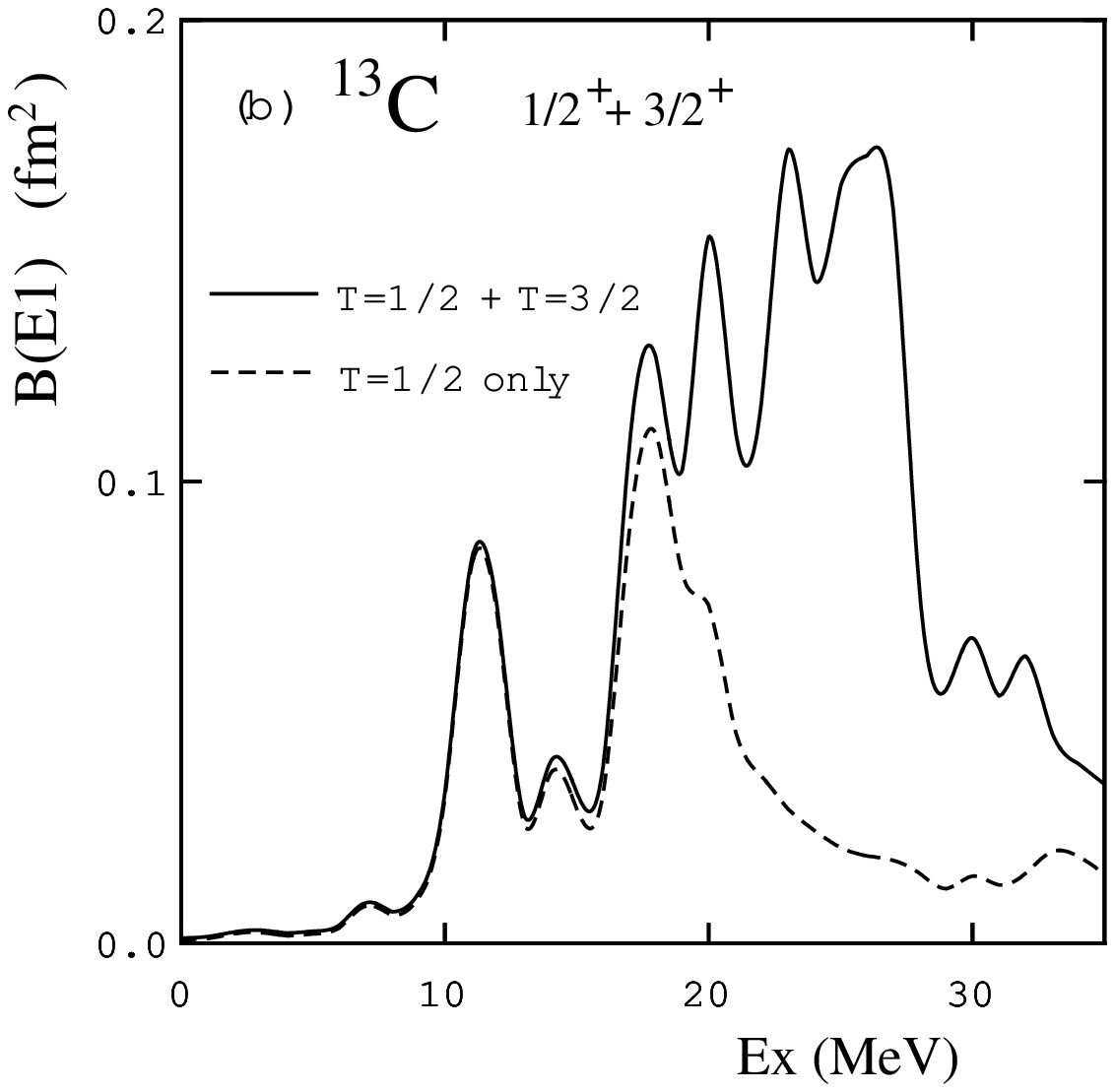,height=3.6in}
\vspace{-10mm}
\hspace{-20mm}
\psfig{figure=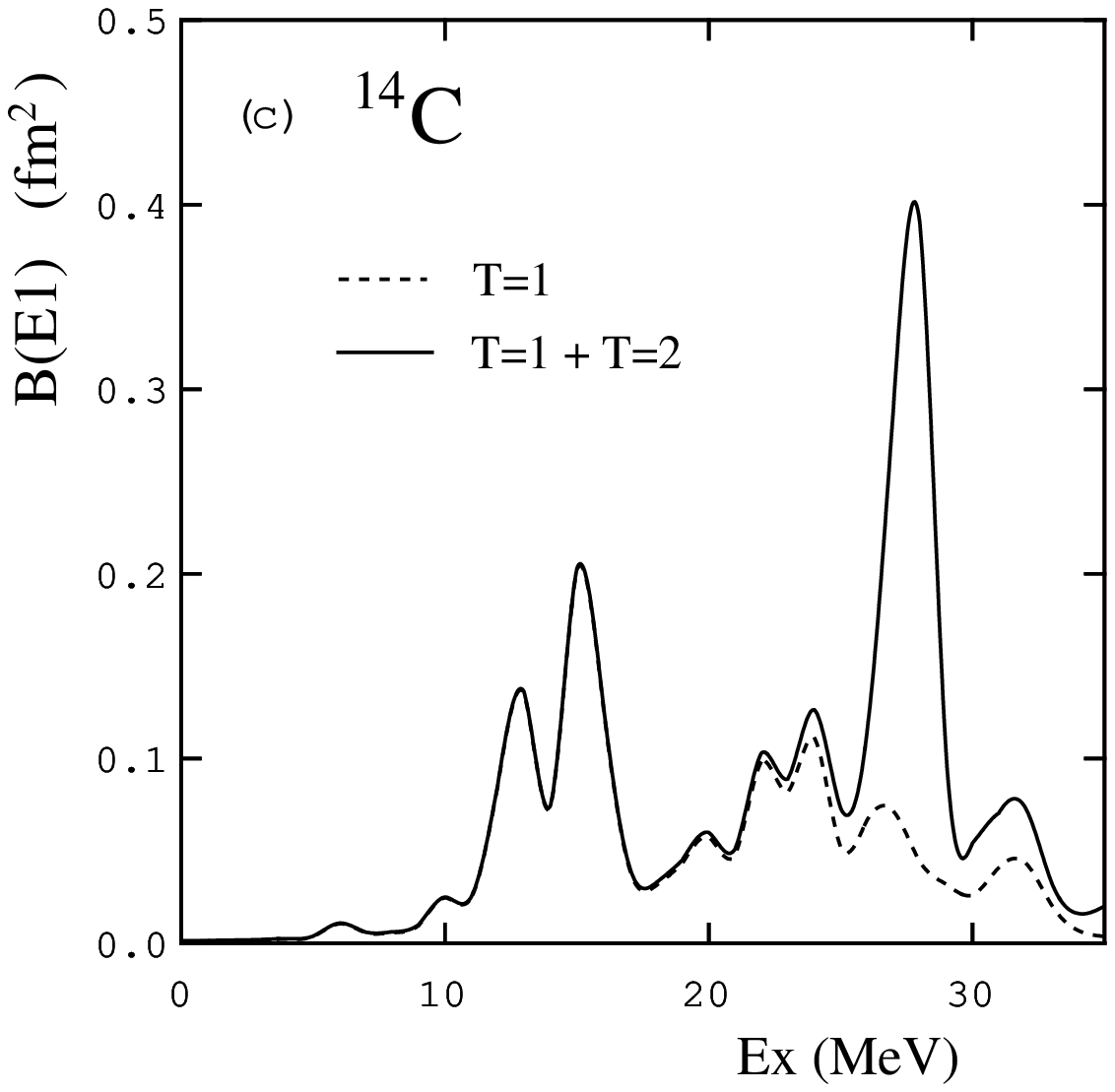,height=3.6in}
\vspace{2.5cm}
\caption{Calculated B(E1) strength for $^{12}$C,  for $^{13}$C and 
 for $^{14}$C with the
use of the WBP10 interaction including up to 3$\hbar\omega$ excitations. 
(a) The solid curve shows the results with (1$+$3)$\hbar\omega$ excitations in
 $^{12}$C, while the dashed curve  gives those with 1 $\hbar\omega$ excitations
only.  (b) The solid curve  includes both  the results of T=1/2 and 3/2 states
 in $^{13}$C, while
the dashed curve  gives the ressults of  T=1/2 states only.  The final states 
are  J$^{\pi}$=1/2$^+$ and 3/2$^+$ states. (c) The solid curve includes both  
the results of T=1 and  T=2 states, while
the dashed curve  gives the results of  T=1 states only.
\label{fig:fig8}}
\end{center}
\end{figure}

\begin{figure}[tbh]
\begin{center}
\vspace{-10mm}
\hspace{-29mm}
\psfig{figure=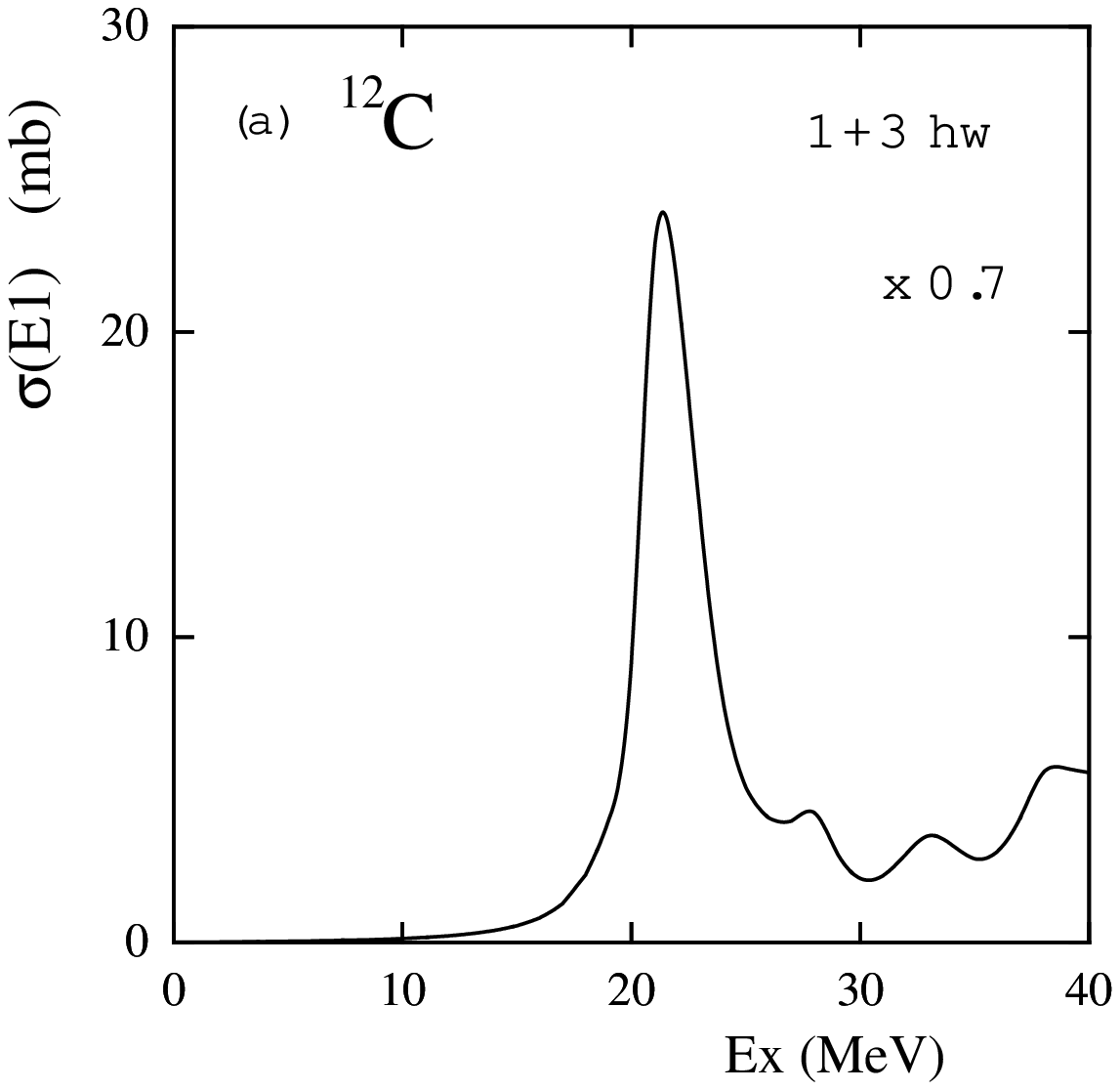,height=3.6in}
\vspace{-10mm}
\hspace{-18mm}
\psfig{figure=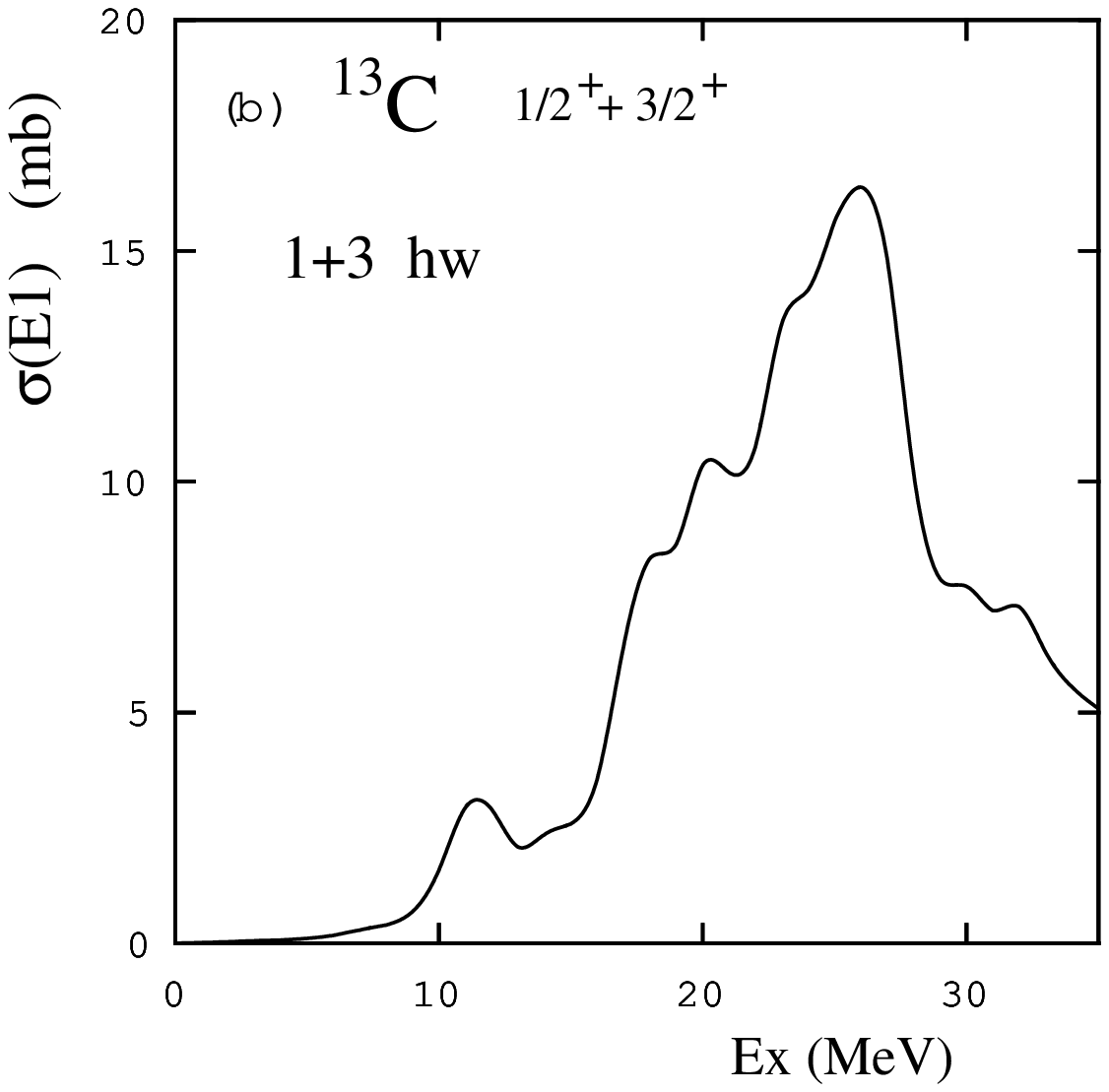,height=3.6in}
\vspace{-10mm}
\hspace{-18mm}
\psfig{figure=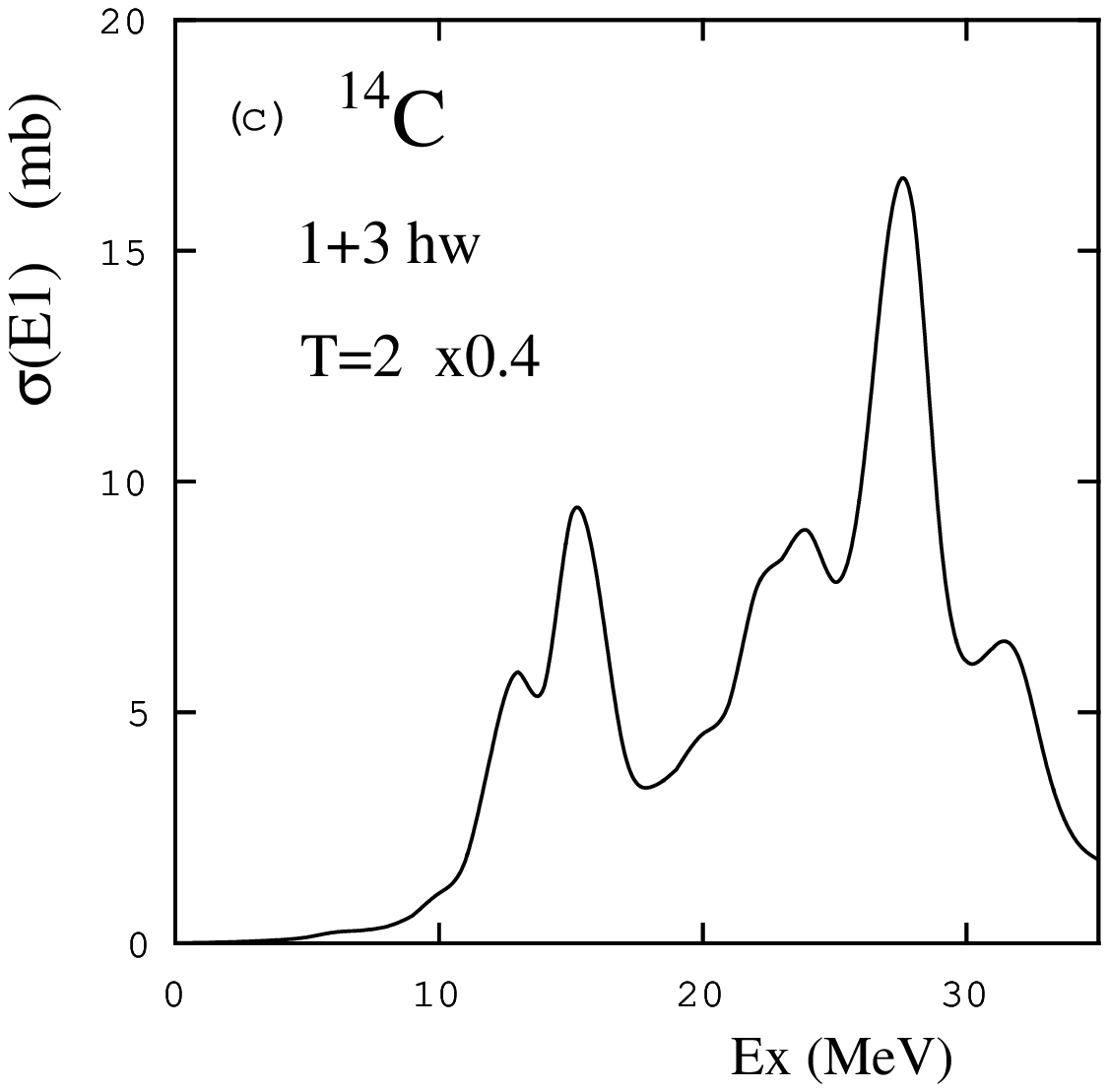,height=3.6in}
\vspace{2.6cm}
\caption{Calculated photoreaction cross sectios for $^{12}$C, $^{13}$C and $^{14}$C. The 
T$_>$ part of giant resonances  is multiplied by quenching factors 
 0.7, 1.0 and 0.4 for $^{12}$C, $^{13}$C and $^{14}$C,
respectively.  The shell model calculations are performed with the
use of the WBP10 interaction including up to 3$\hbar\omega$ excitations. 
\label{fig:fig9}}
\end{center}
\end{figure}

%\begin{figure}[tbh]
\begin{figure}[h]
\begin{center}
\vspace{-10mm}
\hspace{-20mm}
\psfig{figure=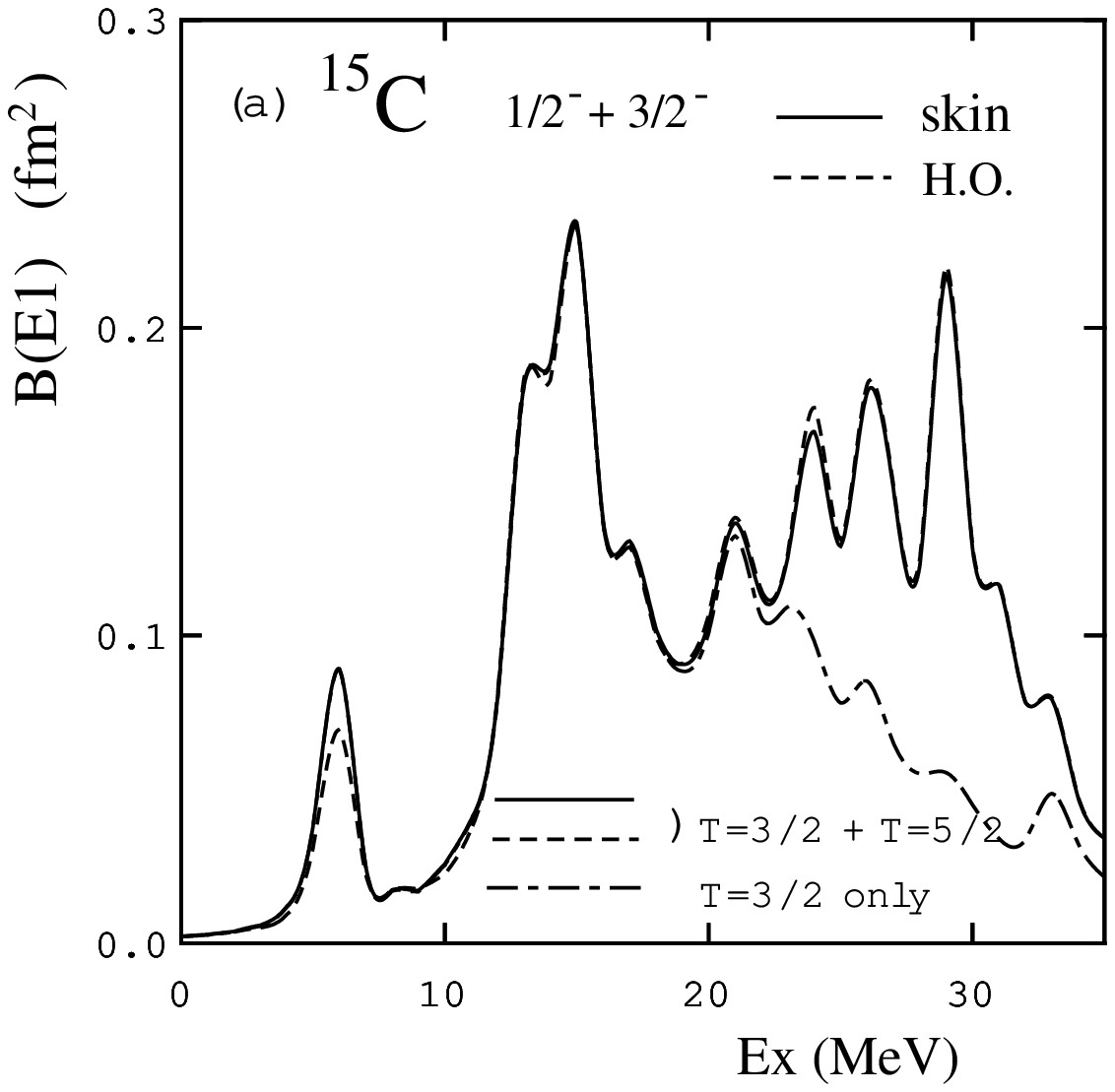,height=3.6in}
\vspace{-10mm}
\hspace{-20mm}
\psfig{figure=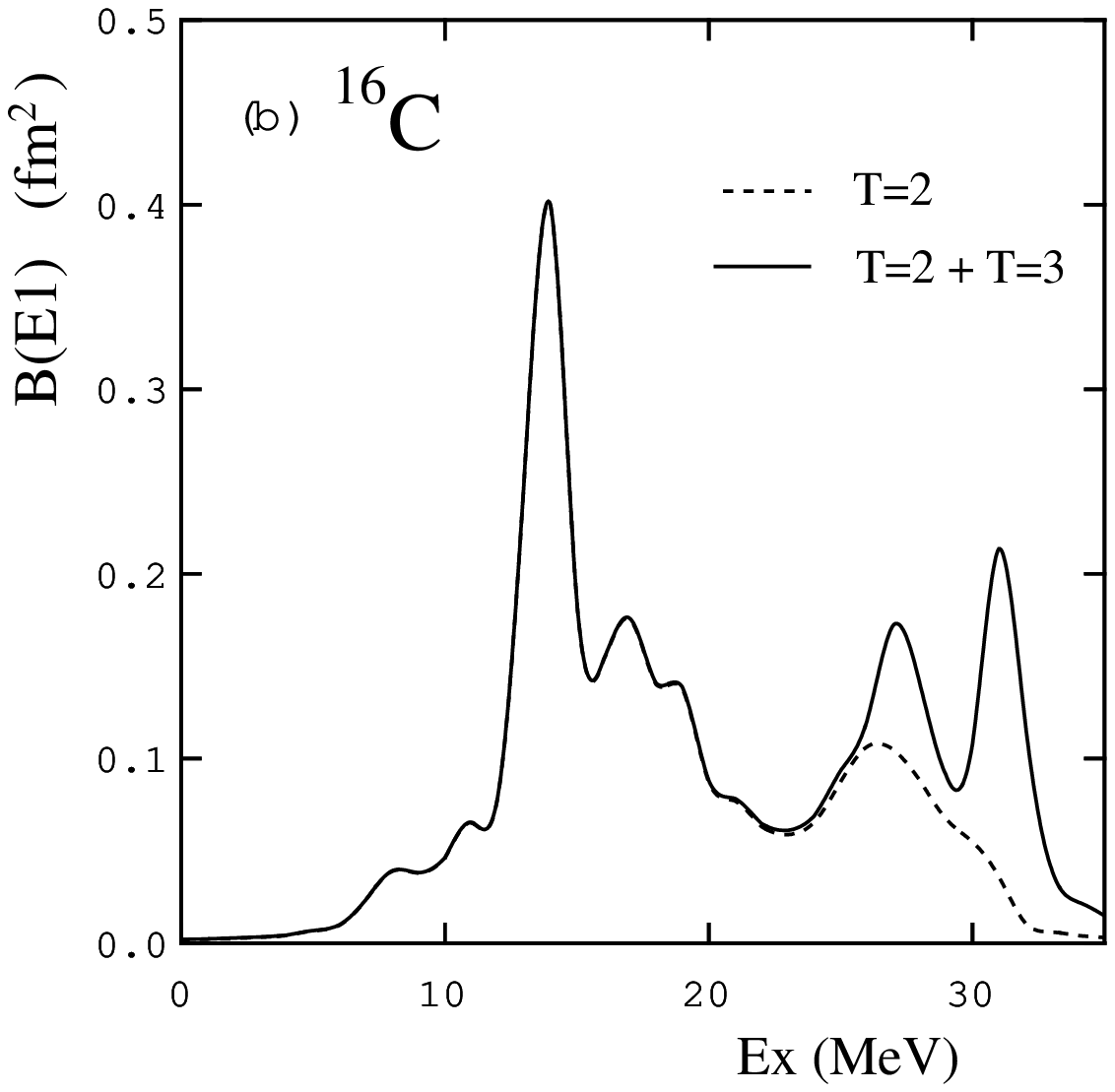,height=3.6in}
\vspace{-10mm}
\hspace{-20mm}
\psfig{figure=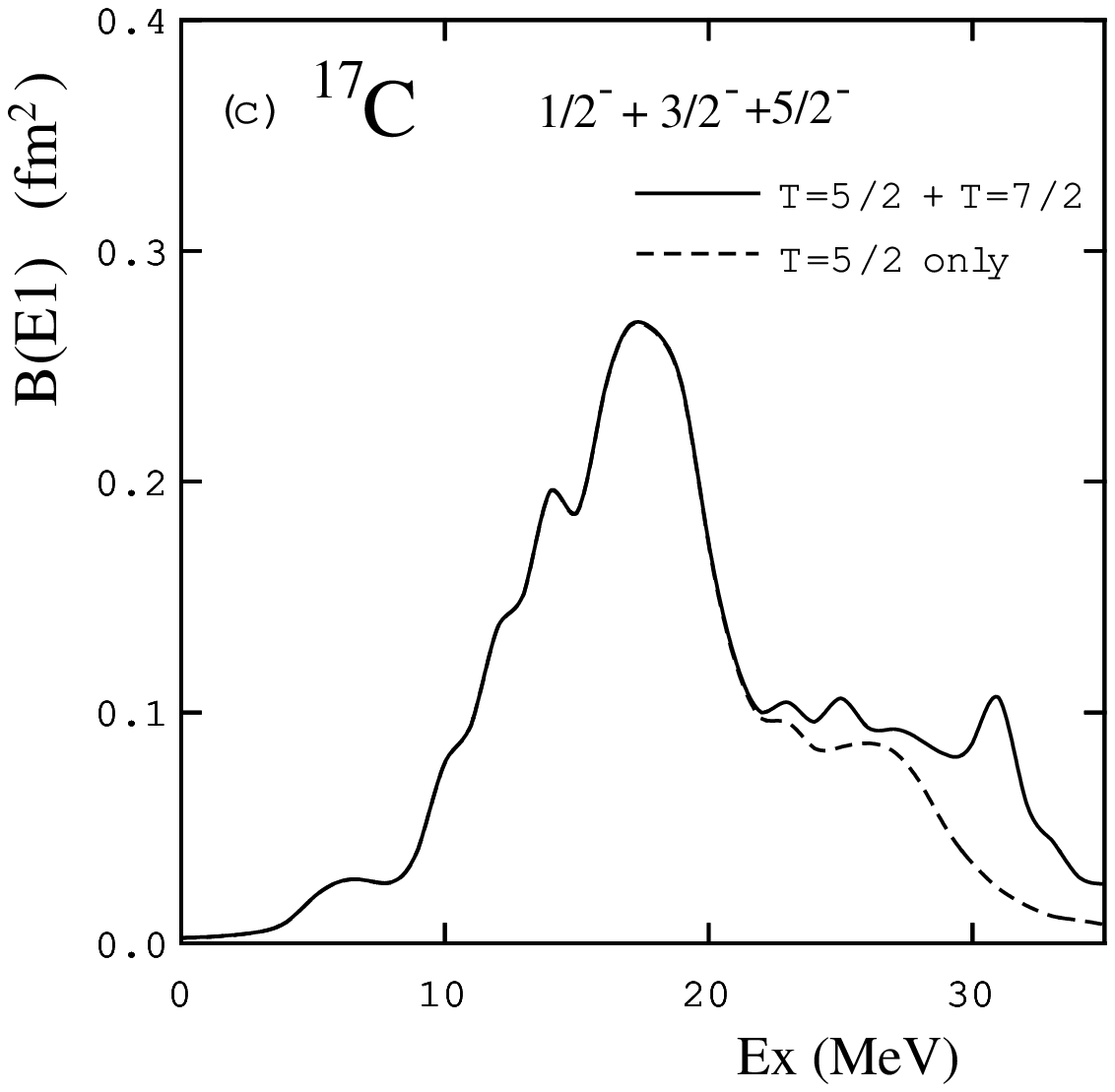,height=3.6in}
\vspace{10mm}
\hspace{-20mm}
\psfig{figure=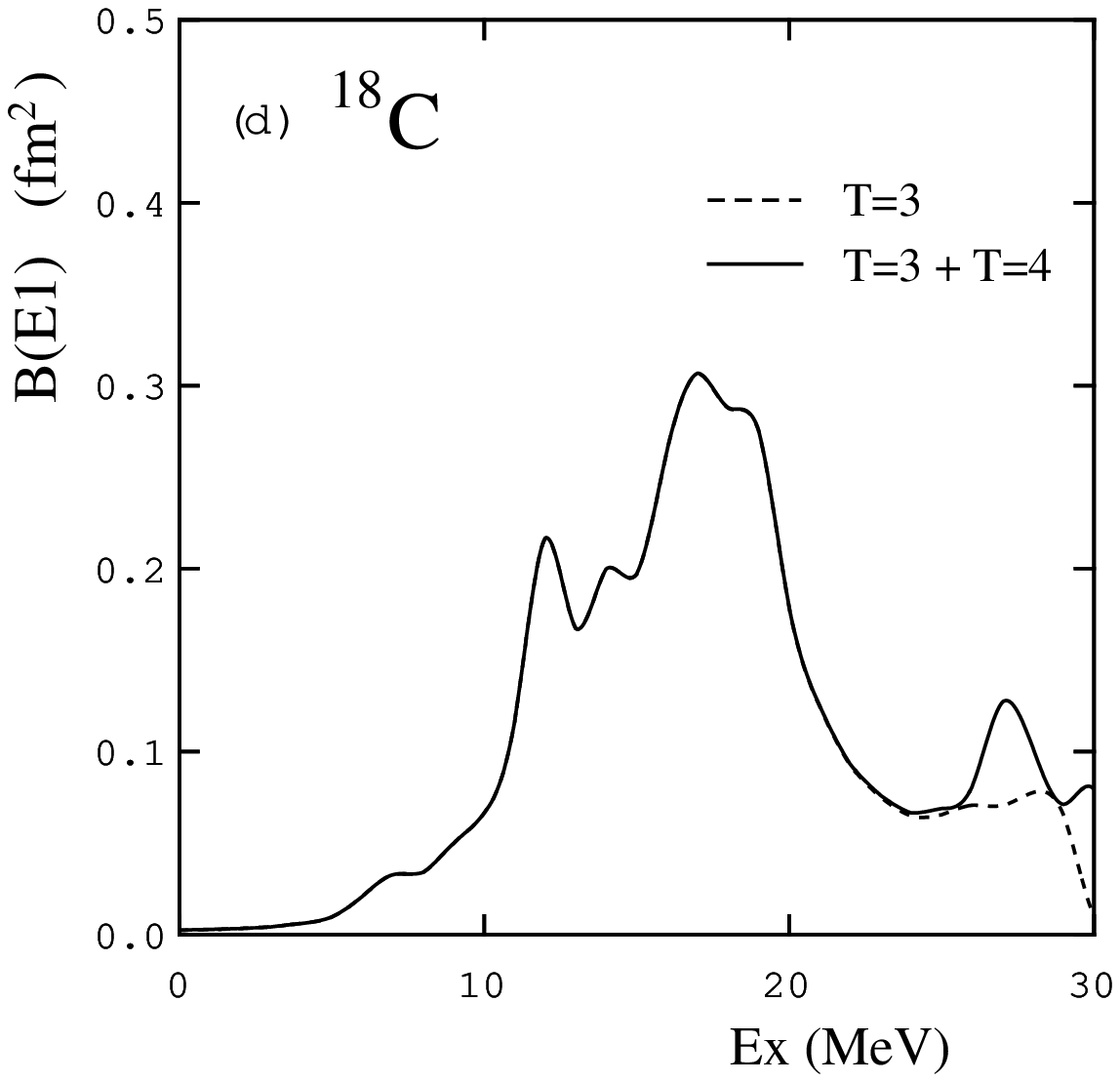,height=3.6in}
\vspace{10mm}
\hspace{-20mm}
\psfig{figure=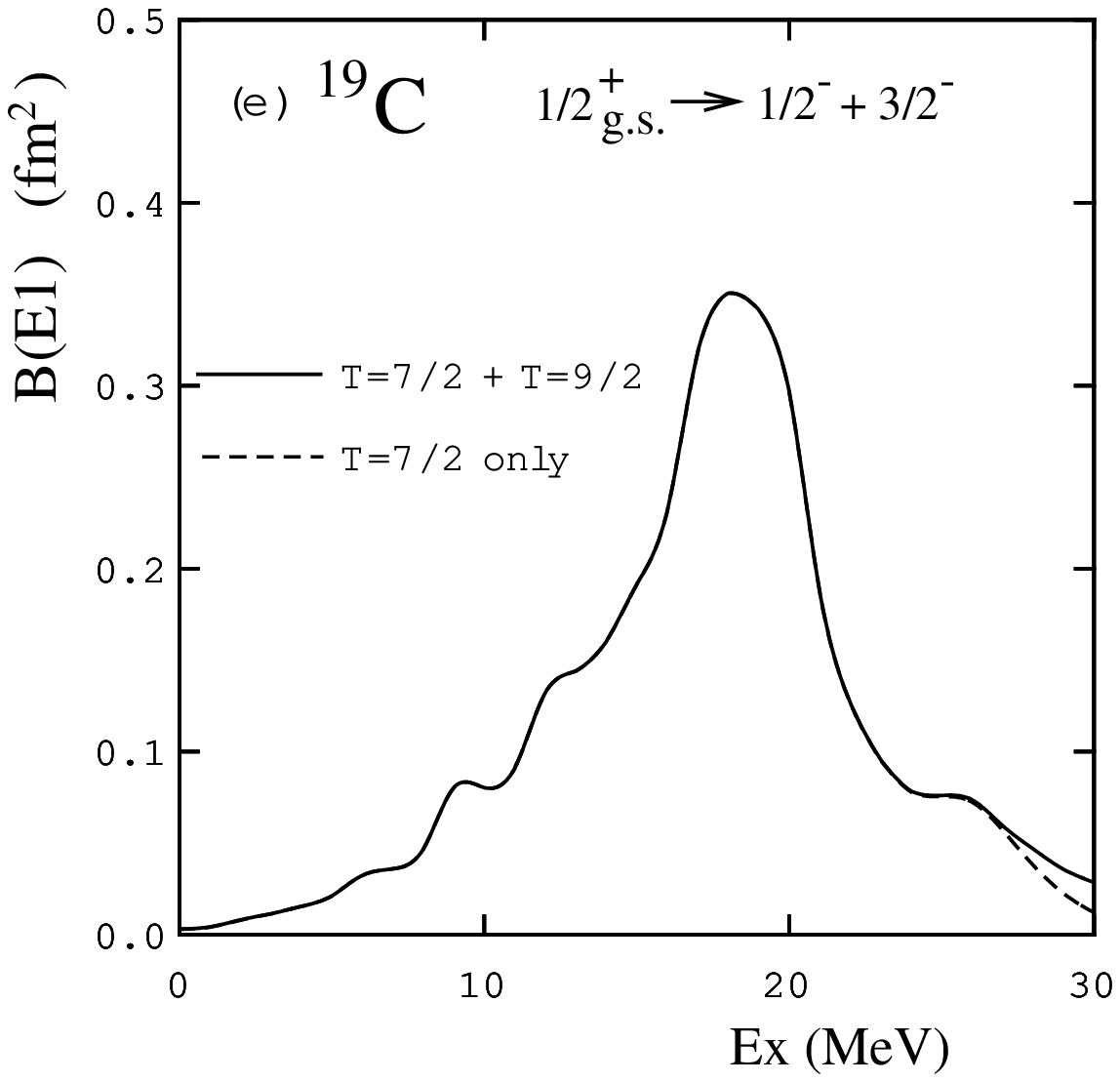,height=3.6in}
\vspace{10mm}
\hspace{-20mm}
\psfig{figure=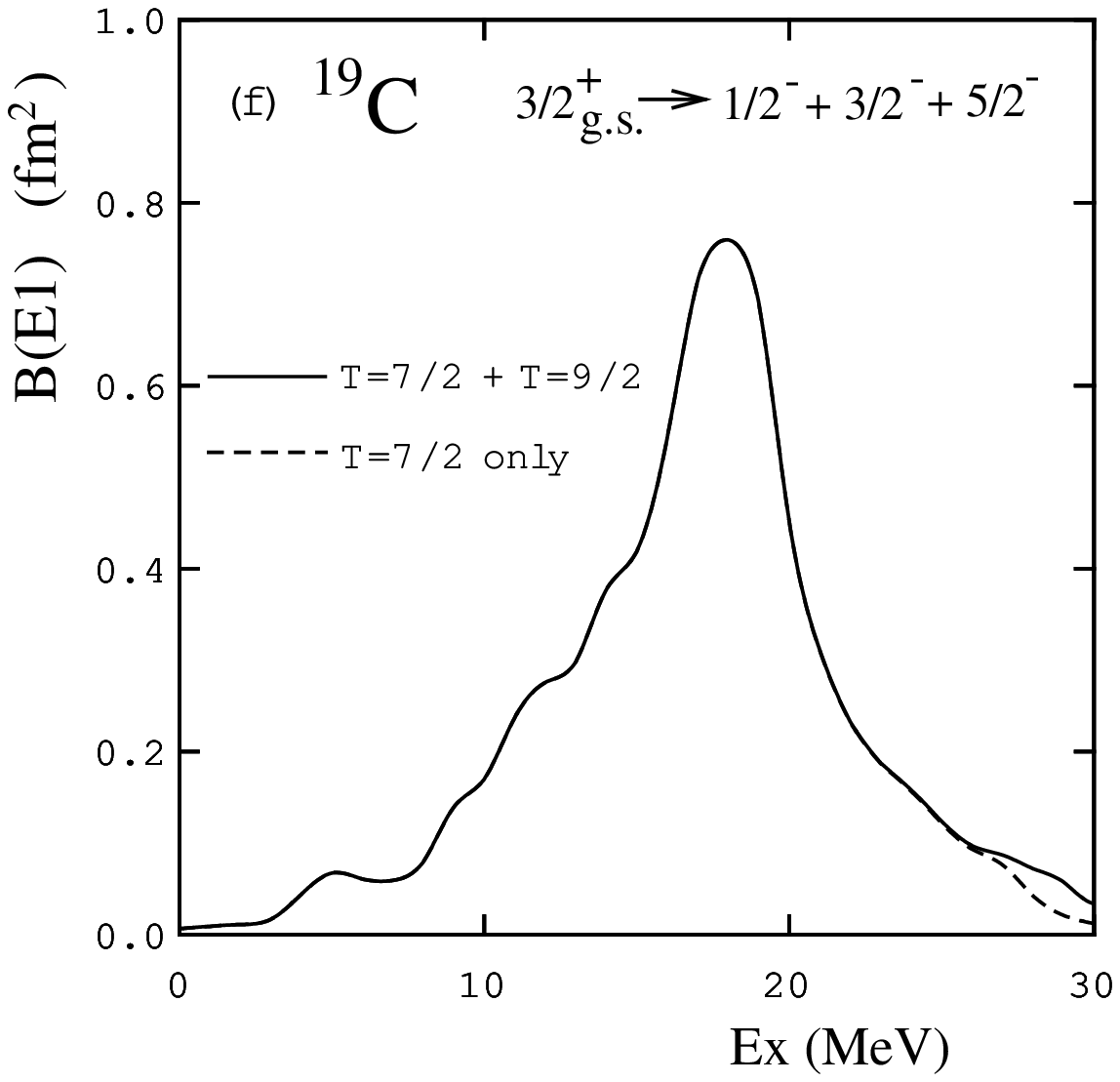,height=3.6in}
\vspace{-2.0cm}
\caption{Calculated B(E1) strength for C-isotopes with N=9$\sim$13.  
with the use
of the WBP10 interaction and the configurations
 of 1$\hbar\omega$ excitations.  The solid curve includes the results
 of  both T$_<$  and  T$_>$ states, while the dashed  curve includes only 
 those of  T=$_<$   in all figures except in $^{15}$C.  (a)  $^{15}$C; 
The solid curve includes the results
 of  both T$_<$  and  T$_>$ states , while the dashed-dotted curve includes only 
 those of  T=$_<$ states .  
The dashed curve includes the effect of neutron 
skin for  both  T$_<$  and  T$_>$   states.  (b)  $^{16}$C.  (c) $^{17}$C.
(d) $^{18}$C. (e)  $^{19}$C; The ground state is taken to be 1/2$^{+}$. 
(f)  $^{19}$C; The ground state is taken to be 3/2$^{+}$.
\label{fig:fig10}}
\end{center}
\end{figure}

\end{document}